\documentclass[aps,article,nofootinbib,groupedaddress]{revtex4}
\usepackage{graphicx}
\usepackage{amsmath,amssymb}
\usepackage{bm}
\usepackage{hyperref}
\usepackage[usenames]{color}

\hypersetup{
    colorlinks=true,
    linkcolor=red,
    citecolor=blue,
}

\def\be{\begin{equation}}
\def\ee{\end{equation}}
\def\ba{\begin{eqnarray}}
\def\ea{\end{eqnarray}}

\def\bmk{\bm k}
\def\bmp{\bm p}

\def\bmq{\bm Q}
\def\gd{\gamma_D}
\def\go{\gamma_\omega}
\def\gb{\gamma_B}
\def\gdc{\gamma_{ D c}}
\def\goc{\gamma_{\omega c}}
\def\gbc{\gamma_{Bc}}

\begin{document}

\title{
Chiral Effects and Cosmic Magnetic Fields
}

\date{\today}

\author{Hiroyuki Tashiro, Tanmay Vachaspati}
\affiliation{ 
Physics Department, Arizona State University, Tempe, AZ 85287, USA.
}

\author{Alexander Vilenkin}
\affiliation{ 
Institute of Cosmology, Department of Physics \& Astronomy, 
212 College Avenue, Tufts University, 
Medford, MA 02155, USA.
}

\begin{abstract}
In the presence of cosmic chiral asymmetry, chiral-vorticity and 
chiral-magnetic effects can play an important role in the generation 
and evolution of magnetic fields in the early universe. We include
these chiral effects in the magnetic field equations and find solutions
under simplifying assumptions. Our numerical and analytical
results show the presence of an attractor solution in which chiral 
effects produce a strong, narrow, Gaussian peak in the magnetic 
spectrum and the magnetic field becomes maximally helical. 
The peak in the spectrum shifts to longer length scales and becomes 
sharper with evolution. We also find that the dynamics may become
non-linear for certain parameters, pointing to the necessity of a 
more complete analysis.
\end{abstract}

\maketitle

\section{Introduction}
\label{intro}

Cosmic relics, such as topological defects and magnetic fields, can be
used as probes of the very early universe and extremely high energy
particle physics. The formation, evolution, and observational signatures
of topological defects have been studied for over three decades, and
constraints have been obtained that have guided particle physics model
building. The formation of magnetic fields during cosmological phase
transitions has also been investigated \cite{Vachaspati:1991nm,
Vachaspati:1994xc,Vachaspati:2008pi}.
Current observational constraints are relatively weak, allowing for 
intergalactic magnetic field strengths at the nano Gauss level
(see for example, \cite{Kosowsky:1996yc,Harari:1996ac,Kosowsky:2004zh,
Kahniashvili:2008hx,Pogosian:2011qv}).

An attractive scenario links the generation of magnetic fields
to the generation of the observed matter-antimatter asymmetry 
\cite{Cornwall:1997ms,Vachaspati:2001nb,Copi:2008he,
Chu:2011tx}. The produced magnetic field carries magnetic helicity
that is directly proportional to the baryon number density
\cite{Cornwall:1997ms,Vachaspati:2001nb}
\begin{equation}
h = - \kappa \frac{n_b}{\alpha},
\label{hnb}
\end{equation}
where $\alpha$ is the fine structure constant, 
$\kappa \approx 0.01$ \cite{Copi:2008he, Chu:2011tx}, and 
\begin{equation}
h = \frac{1}{V} \int_V d^3x ~ {\bf A}\cdot {\bf B},
\label{hdef}
\end{equation}
is the magnetic helicity density and $n_b$ is the baryon number
density. The minus sign in the relation (\ref{hnb}) is a direct
cosmological manifestation of CP violation in particle physics
that also gives preference to matter over antimatter in the universe.
The injection of helical magnetic fields into the plasma can transfer
magnetic field power to larger length scales by the ``inverse cascade'', 
providing hope that even small scale magnetic fields from phase
transitions can grow to astrophysically relevant scales at more recent 
epochs. {\it Helical} magnetic fields can possibly be detected through 
various cosmological observations
\cite{Caprini:2003vc, Kahniashvili:2005xe, Kahniashvili:2005yp}.

\begin{figure}
  \includegraphics[height=0.30\textwidth,angle=0]{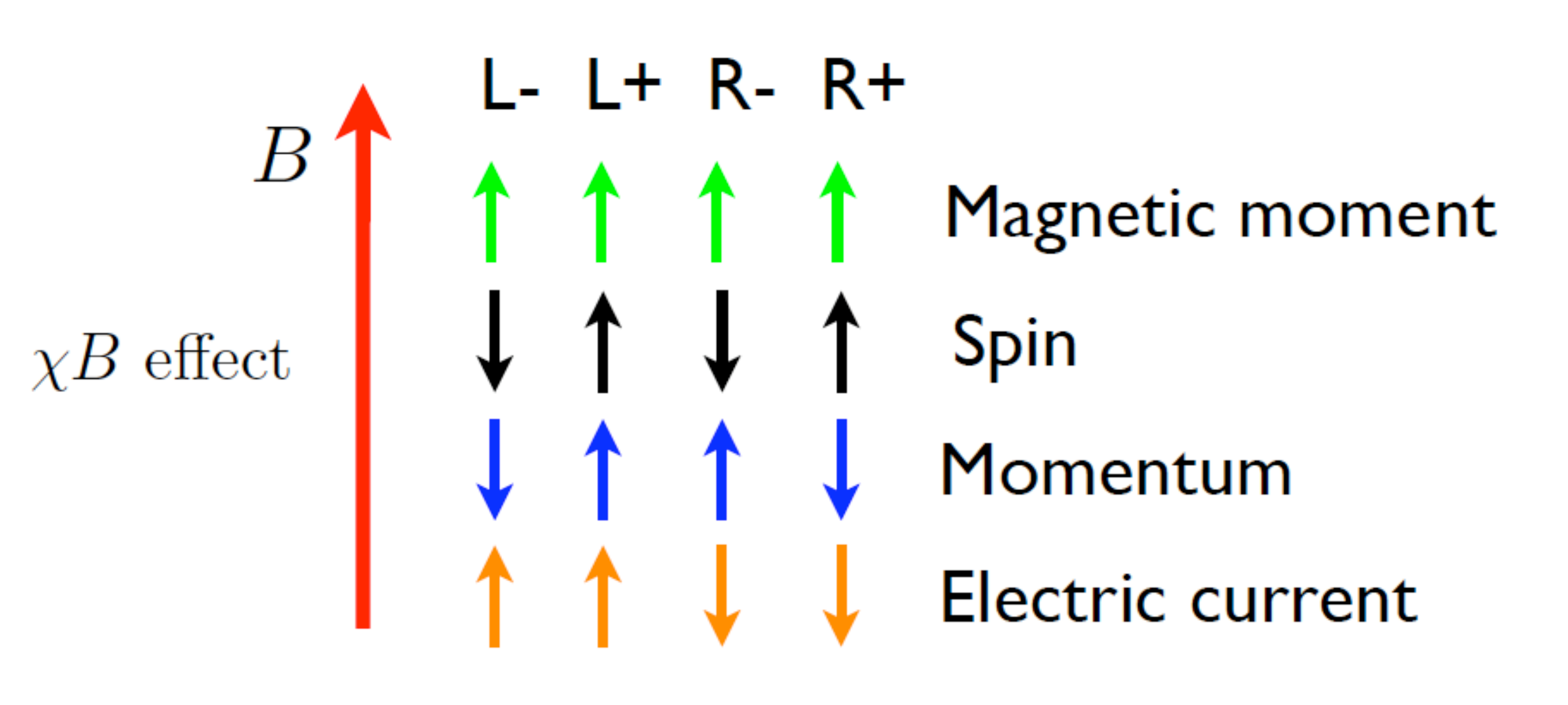}
\caption{
Understanding the $\chi B$ effect. An external magnetic 
field tends to align the magnetic moments of the four electron 
states -- left-right handedness for electron and positron, denoted
in the figure as $L+,L-,R+,R-$ --  which implies the shown 
directionalities of the spin, momenta, and electric current due 
to each state . If the four states are present in unequal numbers, 
net electric current may be induced.
}
\label{chi-B-scheme}
\end{figure}

\begin{figure}
  \includegraphics[height=0.30\textwidth,angle=0]{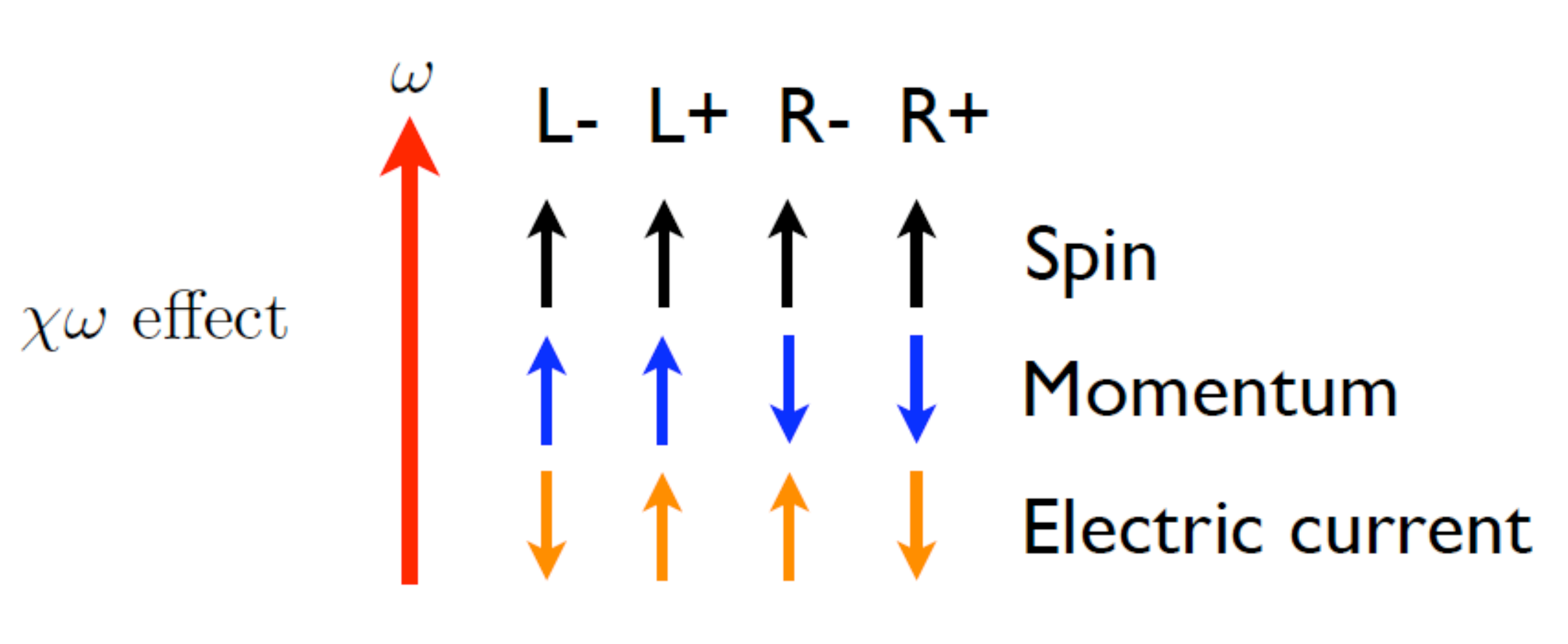}
  \caption{
Understanding the $\chi \omega$ effect. Vortical fluid flow
tends to align the spins of the four electron states which implies 
the shown directionalities of the momenta and electric current due 
to each state . If the four states are present in unequal numbers, 
net electric current may be induced.
}
\label{chi-omega-scheme}
\end{figure}

Two important chiral effects, called the chiral-vorticity 
($\chi \omega$) \cite{Vilenkin:1979ui} and chiral-magnetic 
($\chi B$) \cite{Vilenkin:1980fu} effects, can also play a role
in the early universe \cite{Vilenkin:1982pn} and in 
QCD \cite{Kharzeev2006,Kharzeev:2011vv}.  To understand these effects, 
consider Fig.~\ref{chi-B-scheme} where we show the effect of a magnetic
field on, for example, electrons. 
The magnetic field couples to the magnetic moments and tends to align
them. Depending on the electric charge of the carrier, the spins
are either aligned or anti-aligned with the magnetic moment.
The helicity eigenstates of the fermions then determine the
direction of the momentum of the particles, which in turn gives
the direction of the electric current due to each species. 
Taking the left- and right-handed electrons and positrons as 
the four fermion states, which we assume to be massless, we see 
that the net electric current is
\begin{equation}
J_{\chi B} \propto [n(e^-_L)-n(e^+_R)] - [n(e^-_R)-n(e^+_L)],
\end{equation}
where $n(e^-_L)$ denotes the number density of $e^-_L$ and similarly
for the other particle species. 
In terms of the chemical potentials for left- and right-handed
electrons, the differences within the square brackets are given 
by $\mu_L$ and $\mu_R$ respectively. So 
$J_{\chi\omega} \propto \Delta\mu \equiv \mu_L - \mu_R$. The
calculation in Ref.~\cite{Vilenkin:1980fu} gives
\begin{equation}
{\bm J}_{\chi B} = \frac{e^2}{2\pi^2} \Delta\mu ~{\bm B}.
\label{chiBj}
\end{equation}
Similarly, in Fig.~\ref{chi-omega-scheme}, we explain the
$\chi\omega$ effect, which occurs if the ambient fluid flow
has vorticity (${\bm \omega}$). Spin-orbit coupling tends
to align the spins of the fermions; particle helicity then
aligns the left-handed states but anti-aligns the right-handed
states, which leads to the electric currents as shown. Thus,
in equilibrium,
\begin{equation}
J_{\chi\omega} \propto [n(e^-_L)+n(e^+_R)] - [n(e^-_R)+n(e^+_L)].
\end{equation}
The presence of non-zero $\mu_L$ means
that $n(e^-_L)\ne n(e^+_R)$ and of $\mu_R$ that $n(e^-_R) \ne n(e^+_L)$.
However, if $\mu_L=\mu_R$ then $n(e^-_L)=n(e^-_R)$ and $n(e^+_R)=n(e^+_L)$,
and $J_{\chi\omega}$ vanishes. Also if $\mu_L=-\mu_R$ then 
$n(e^-_L)=n(e^+_L)$ and $n(e^+_R)=n(e^-_R)$, and again $J_{\chi\omega}=0$.
So for $J_{\chi\omega}$ to be non-vanishing, we need 
$\Delta\mu^2 \equiv \mu_L^2 - \mu_R^2 \ne 0$. The exact calculation
in Ref.~\cite{Vilenkin:1979ui} gives
\begin{equation}
{\bm J}_{\chi\omega} = \frac{e}{4\pi^2} \Delta\mu^2 ~{\bm \omega}\ ,
\label{chiomegaj}
\end{equation}
where ${\bm \omega}={\bm \nabla}\times {\bm v}$ is the fluid vorticity.

The above expression for ${\bm J}_{\chi\omega}$ holds when the left-
and right-handed particles and antiparticles are in thermal equilibrium
at the same temperature. If some of the species are at different temperatures 
there is an additional contribution per species to ${\bm J}_{\chi\omega}$ 
proportional to $e T^2 {\bm \omega}$ where $T$ is the temperature of the 
particular species \cite{Vilenkin:1979ui}. 
We will not consider this situation in the present
paper, though it may be important for the contribution of left- and
right-handed particles, especially neutrinos, to the hypercharge current 
in the epoch before electroweak symmetry breaking.

The $\chi B$ and $\chi\omega$ effects can only lead to a non-zero electric
current if there is a disbalance between left- and right-handed particles, that
is, $\Delta\mu \ne 0$.   Such a disbalance can arise in the early universe from
out-of-equilibrium $P$-violating decays of massive particles.  For example, it
could arise due to inflaton decay at the end of inflation.  The resulting
values of $\Delta\mu$ are not generally suppressed by any small couplings, so
large values like $\Delta\mu/T \sim 0.1 - 1$ can easily be achieved 
\cite{KolbTurner,Joyce:1997uy}.  A right-left asymmetry is also likely to be
created during leptogenesis, if it occurs at some energy scale much higher than
the electroweak scale.  In particular, an asymmetry between left- and
right-handed electrons will be created because they carry different charges
under the electroweak symmetry group $SU(2)_L \times U(1)_Y$.  The
chirality-flipping processes due to nonzero electron mass are suppressed at
high temperatures,  and this asymmetry is preserved until the temperature drops
below $T_f \sim 80~{\rm TeV}$ \cite{Campbell:1992jd}.
If vorticity develops at any temperature higher than $T_f$,
the $\chi\omega$ effect will operate and create a magnetic
field. Since the electroweak symmetry is unbroken, this magnetic
field will be the $U(1)$ hypercharge magnetic field, and will
get converted to electromagnetic magnetic field after the
electroweak phase transition.

It is worth noting that chiral effects induce kinetic and magnetic 
helicities even if these are not present initially. For the 
$\chi\omega$ effect, Fig.~\ref{chi-omega-scheme} shows that
momenta tend to be aligned along ${\bm \omega}$, which means that
${\bm v}\cdot {\bm \omega}\ne 0$ and so kinetic helicity is
induced. In the case of the $\chi B$ effect, if the magnetic 
field is helical, then the induced current, hence velocity, is 
along the magnetic field and the fluid flow carries kinetic helicity. 
This fact may be of interest if the kinetic helicity survives
until recombination because then it can produce parity odd
temperature-polarization correlations \cite{Pogosian:2001np}.

Once a magnetic field is generated, its subsequent evolution may also be
influenced by chiral effects.  In particular, Joyce and Shaposhnikov
\cite{Joyce:1997uy} have shown that $\chi B$ currents can induce an exponential
growth of the magnetic field on sufficiently large length scales.
A helical magnetic field, in turn, back-reacts on the evolution of the
left-right disbalance $\Delta\mu$.   This is described by the chiral anomaly
equation, which relates changes in $\Delta\mu$ to 
changes in magnetic helicity \cite{Joyce:1997uy},
\begin{equation}
\frac{d(\Delta\mu)}{dt} = - \frac{c_\Delta \alpha}{T^2} \frac{dh}{dt}
 - \Gamma_F \Delta \mu.
\label{Dmuevoln}
\end{equation}
Here, $\Gamma_F$ is the chirality-flipping rate and 
$c_\Delta \sim 1$, $\alpha$ is the fine structure constant, and $T$
is the temperature at time $t$.  Boyarsky et al \cite{Boyarsky:2011uy} have
shown that as a result the left-right disbalance can survive down to 
$T\sim 10$~MeV if helical magnetic fields are present, for example due to 
baryogenesis.

The plasma equations in the radiation-baryon single fluid approximation are
\begin{eqnarray}
\partial_t\rho + {4 \over 3}
 {\bm \nabla}\cdot (\rho {\bm v}) &=&0 ,
\\
{4 \over 3} \partial_t (\rho {\bm v})  - {4 \over 3} \rho{\bm v} \times ({\bm
 \nabla} \times {\bm v}) &=&
 {\bm J}\times {\bm B} -{\bm \nabla}p, \\
{\bm \nabla}\times {\bm B} &=& {\bm J} ,
\label{Max1}\\
{\bm \nabla}\times {\bm E} &=& - \partial_t {\bm B},
\label{Max2}
\end{eqnarray}
where we have used natural units, $\hbar=1=c$. 
The fluid density is $\rho$, the pressure is $p$.
The displacement current has been ignored, as is done in 
magnetohydrodynamics (MHD) when the flow velocities are small 
compared to the speed of light. Since we are mainly focusing on
chiral effects, we have simplified the equations by ignoring
the injection of magnetic fields by external sources such as 
sphalerons \cite{Vachaspati:2001nb},
and the dissipative effects of viscosity. Cosmological
expansion can be included in the Maxwell equations by going to 
conformal coordinates, as we will do in Sec.~\ref{sec:diffevoln}.

The electric current is given by the sum of the Ohmic and chiral
components
\begin{equation}
{\bm J} = {\bm J}_{\rm Ohm}+ {\bm J}_{\chi\omega} + {\bm J}_{\chi B}.
\end{equation}
The Ohmic component is given by
\begin{equation}
{\bm J}_{\rm Ohm} = \sigma ({\bm E}+{\bm v}\times {\bm B}) .
\end{equation}
where $\sigma$ is the electrical conductivity of the plasma. The
chiral components, ${\bm J}_{\chi B}$ and ${\bm J}_{\chi\omega}$,
are given in Eqs.~(\ref{chiBj}) and (\ref{chiomegaj}), together
with the chiral anomaly equation in the form of Eq.~(\ref{Dmuevoln}).

In Eq.~(\ref{Max2}) we can replace ${\bm E}$ by the currents and
with a little algebra we find
\begin{equation}
\partial_t {\bm B} = {\bm \nabla}\times ({\bm v}\times {\bm B})
 + \gd \nabla^2 {\bm B} + \go {\bm \nabla}\times {\bm \omega} + 
    \gb {\bm \nabla}\times {\bm B},
\label{mhdeq}
\end{equation}
where
\begin{equation}
\gd = \frac{1}{\sigma} \ , \ \ 
\go = \frac{e \Delta \mu^2}{4\pi^2 \sigma} \ , \ \ 
\gb = \frac{e^2 \Delta \mu}{2\pi^2 \sigma} 
\label{eqabc}
\end{equation}
The terms on the right-hand side of (\ref{mhdeq}) will be referred
to as the advection, diffusion, $\chi\omega$, and $\chi B$ terms
respectively.
The above set of equations exhibits all the complexities of MHD
plus those arising from the chiral anomaly.  

It is helpful to compare order of magnitudes of the various
terms on the right-hand side of Eq.~(\ref{mhdeq}). If $L$
denotes a length scale of interest, the advection, diffusion,
$\chi\omega$ and $\chi B$ terms are estimated as
\begin{equation}
{\bm \nabla}\times ({\bm v}\times {\bm B}) \sim \frac{vB}{L} \ , \ \
\gd \nabla^2 {\bm B} \sim \frac{B}{\sigma L^2} \ , \ \ 
\go {\bm \nabla}\times {\bm \omega} \sim 
        \frac{e \Delta \mu^2 v}{\sigma L^2} \ , \ \ 
\gb {\bm \nabla}\times {\bm B} \sim 
                \frac{e^2 \Delta \mu B}{\sigma L} \ .
\label{eq:terms}
\end{equation}

Assuming that $\mu_L \sim \mu_R \sim T$ and using $\sigma \sim T/e^2$
\cite{Turner:1987bw,Baym:1997gq}, the estimates for the four terms become
$vB/L$, $e^2 B/(T L^2)$, $e^3 vT/L^2$, and $e^4 B/L$ respectively. 
Now for small flow velocities, $v \ll \max \{e^2/(T L), ~e^4\}$,
or small magnetic fields, $B \ll e^3 T/L$, the advection term is
subdominant and can be ignored.
The second case we consider is when the length scale is large and
the magnetic field is small: 
$L \gg \min \{e^2/(v T), ~ 1/(e^2 T)\}$, $B \ll e v T^2$. 
Then the diffusion term can be neglected. 

We solve the evolution equations in an expanding spacetime, 
first numerically in Sec.~\ref{sec:numerics} and then analytically 
in Sec.~\ref{analytics}. Throughout this analysis 
we adopt some simplifications.  First of all, 
we do not attempt to solve the fluid dynamics equations and represent 
the velocity field by a mode distribution with a Kolmogorov spectrum 
and fixed phases.  (A more realistic representation of
turbulent dynamics would include time variation of the phases.)  
We also assume negligible advection; the consistency of the latter assumption
is discussed 
in Sec.~\ref{analytics}.  Our conclusions are summarized and discussed in
Sec.~\ref{conclusions}.
In particular, we argue that under certain conditions the magnitude and the
spectrum of the magnetic fields produced by our mechanism are not sensitive to
the details of the turbulent velocity flow.  Our estimates should then be
valid, despite the oversimplified treatment of turbulent dynamics.  In
Appendix~\ref{app:steady} we also consider 
steady-state solutions, {\it i.e.} with $\partial_t {\bm B}=0$, in 
the regimes of negligible advection and of negligible diffusion in 
Minkowski spacetime.

\section{Magnetic Field Generation and Evolution}
\label{sec:diffevoln}

As discussed at the end of Sec.~\ref{intro}, 
there is a range of conditions within which the advection term
may be dropped.  Without the advection term, Eq.~(\ref{mhdeq}) is linear
in ${\bm v}$ and ${\bm B}$, and a mode expansion converts the equation
into a set of ordinary differential equations. However, the dynamics is
still highly non-trivial because of the mode coupling that arises due to
the evolution of $\Delta\mu$, as the magnetic helicity on the right-hand
side of Eq.~(\ref{Dmuevoln}) is an integral over all modes.  Here we shall 
assume that advection is negligible.  The validity of this assumption will 
be discussed at the end of Sec.~\ref{analytics}.

A complete analysis, which is beyond the scope of the present paper,
would need to include the evolution of the velocity field which is 
governed by the Navier-Stokes equation. 
Here we will consider the simpler 
case of incompressible, turbulent flows, with a specified distribution
of velocities. These assumptions are
valid when the magnetic field energy density is much smaller 
than the kinetic energy of the fluid and we can ignore the
backreaction of the magnetic field on the fluid flow.

In a cosmological setting, we assume a flat Robertson-Walker metric
\begin{equation}
 ds^2 = R^2(\eta) (-d \eta^2 + \delta_{ij} dx^i dx^j), 
\end{equation}
where $R(\eta)$ is the scale factor. It is convenient to choose
$R(\eta )$ to have dimensions of length, and $\eta,~ x^i$ to
be dimensionless. The scale factor is related to cosmic temperature
by $R= 1/T$. In the radiation dominated epoch, we also define the 
conformal time as $\eta = M_*/T$, where 
$M_* = \sqrt{90 / 8 \pi^3 g_*} M_{\rm P}$
and $g_* \sim 100$ is the effective number of relativistic degrees of freedom.
With this normalization, we find $\eta \sim 0.1$ when $T\sim M_{\rm P}$ 
and, for example, $\eta \sim 10^{27}$ at matter-radiation equality when
$T\sim 1~{\rm eV}$.
We also define comoving variables
\begin{equation}
B_c =R^2(\eta) B(\eta), \quad \Delta \mu_c = R(\eta) \Delta \mu ,
\end{equation}
and the comoving electrical conductivity is given by \cite{Baym:1997gq}, 
\begin{equation}
 \sigma_c =R(\eta) \sigma  =70.
\label{sigmavalue}
\end{equation}

In these comoving variables, the magnetic field evolution Eq.~(\ref{mhdeq})
takes the form
\begin{equation}
 \partial _\eta {\bm B}_c  = \nabla_c  \times ({\bm v_c} \times {\bm B}_c)
+ \gdc \nabla_c ^2 {\bm B}_c + \goc \nabla_c \times (\nabla_c \times {\bm v_c})
+ \gbc \nabla_c \times {\bm B}_c,
\label{eq:mhdcomoving}
\end{equation}
where
$\gdc=1/\sigma_c$, $\goc= e \Delta \mu_c ^2 /4 \pi^2 \sigma_c$,
$\gbc=e^2 \Delta \mu_c/ 2\pi^2 \sigma_c$, $\nabla_c$ is 
differentiation with respect to the spatial metric $\delta_{ij}$ 
and $v_c$ is the comoving velocity. Note that ${\bm v}_c$ = ${\bm v}$ 
by definition.

From Eq.~(\ref{Dmuevoln}), the evolution of the comoving chemical 
potential is given by
\begin{equation}
 {d \Delta \mu_c  \over d\eta} = - c_\Delta \alpha
  \partial _\eta h_c - \Gamma_F \Delta \mu_c,
\label{eq:mu-evo1}
\end{equation}
where $h_c$ is $R^3 h$ and $\Gamma_F$ is the chirality-flipping rate
that we can ignore in the early universe $T > 80~{\rm TeV}$
\cite{Campbell:1992jd}. In what follows we will always work with
comoving quantities and will omit the subscript $c$ for convenience.

\subsection{Decomposition into modes}
\label{sec:decomp}

In order to solve for the evolution of the magnetic field with a chemical 
potential, we decompose the vector fields in the modes, ${\bm Q}^{\pm} _i$, 
which are divergence-free eigenfunctions of the Laplacian operator in comoving 
coordinates,
\begin{equation}
 {\bm Q}^{\pm}  (\bmk) = \frac{{\bm e_1(\bmk) } \pm i {\bm e_2(\bmk)}}{\sqrt{2}}
  \exp({ i {\bm k \cdot \bm x}}),
\end{equation}
where ${\bm e}_3 = {\bm k}/k$ and $({\bm e}_1,{\bm e}_2,{\bm e}_3)$ form
a right-handed, orthonormal triad of unit vectors. Then,
$\nabla \cdot \bmq^{\pm} =0$ and $\nabla \times \bmq^{\pm} =\pm
k \bmq^\pm$, and we also take $Q^{\pm*} (-{\bm k})=Q^\pm(+{\bm k})$. 

The velocity field of the incompressible (${\bm \nabla}\cdot {\bm v}=0$)
fluid is now decomposed in modes
\begin{equation}
 {\bm v}(\eta, {\bm x} ) = \int {d^3 k  \over (2 \pi)^3} 
       \left [ {\tilde v}^+ (\eta,\bmk) \bmq^+(\bmk) +
                    {\tilde v}^-(\eta,\bmk) \bmq^-(\bmk) \right ].
\label{vdecomp}
\end{equation}
The fluid kinetic energy density is given by
\begin{equation}
{ \rho_r \over 2} \langle |{\bm v}(\eta,{\bm x})|^2 \rangle
          \equiv \frac{\rho_r}{2} \int d \log k~ E_V(\eta, k ) 
     = \rho_r \int {k^2 d k  \over (2 \pi)^2} \left [
      |v^+ (\eta,k)|^2  + |v^- (\eta,k)|^2 \right ] ,
\label{fluidkin}
\end{equation}
where $\rho_r$ is the radiation density, and we have taken statistically
isotropic correlators, 
\begin{eqnarray}
\langle {\tilde v}^{\pm*}(\eta, \bmk){\tilde v}^\pm(\eta,\bmp) \rangle &=& 
    |v^\pm(\eta,k)|^2 ~ (2\pi)^3 \delta^{(3)}(\bmk - \bmp) \\
\langle {\tilde v}^{+*}(\eta,\bmk){\tilde v}^-(\eta,\bmp) \rangle &=& 
\langle {\tilde v}^{-*}(\eta,\bmk){\tilde v}^+(\eta,\bmp) \rangle = 0  .
\end{eqnarray}
Proceeding in the same way, the fluid helicity is
\begin{equation}
\langle {\bm v} \cdot {\bm \omega} \rangle
 \equiv \int d \log k~ H_V(\eta, k ) = \int {k^3 d k  \over 2 \pi^2} 
         \left [ |v^+ (\eta,k)|^2 - |v^- (\eta,k)|^2 \right ] .
\end{equation}

The magnetic field is similarly decomposed 
\begin{equation}
{\bm B}(\eta, {\bm x} ) = \int {d^3 k  \over (2 \pi)^3} \left [
 {\tilde B}^+ (\eta,\bmk) \bmq^+(\bmk)+{\tilde B}^-(\eta,\bmk) \bmq^-(\bmk) 
              \right ] .
\end{equation}
where the scalar mode is absent because $\nabla \cdot {\bm B} =0$.
The ensemble average of the magnetic field energy density is
\begin{equation}
\frac{1}{2} \langle |{\bm B}(\eta, {\bm x})|^2 \rangle
 \equiv \int d \log k~  E_B(\eta,k ) = \int {k^2 d k  \over (2 \pi)^2} 
      \left [ |B^+ (\eta,k)|^2  + |B^- (\eta, k)|^2 \right ].
\label{EB(k)}
\end{equation}
where, just as for the velocity field,
\begin{eqnarray}
\langle {\tilde B}^{\pm*}(\eta,\bmk){\tilde B}^\pm(\eta,\bmp) \rangle &=& 
      |B^\pm(\eta,k)|^2 ~ (2\pi)^3 \delta^{(3)}(\bmk - \bmp) \\
\langle {\tilde B}^{+*}(\eta,\bmk){\tilde B}^-(\eta,\bmp) \rangle &=& 
\langle {\tilde B}^{-*}(\eta,\bmk){\tilde B}^+(\eta,\bmp) \rangle = 0  .
\end{eqnarray}
The magnetic field helicity density is
\begin{equation}
\langle {\bm A} \cdot {\bm B}  \rangle
 \equiv \int d \log k~ H_B(\eta,k ) =
\int {k d k  \over 2 \pi^2} \left [ 
         |B^+ (\eta,k)|^2  -  |B^- (\eta,k)|^2 \right ].
\label{eq:helicitymodes}
\end{equation}
and from Eq.~(\ref{hdef}) we have
\begin{equation}
\langle h \rangle = \langle {\bm A} \cdot {\bm B}  \rangle .
\end{equation}

The MHD equation, Eq.~(\ref{eq:mhdcomoving}), without the advection
term, can be decomposed into equations for the modes ${\tilde B}^\pm$
\begin{eqnarray}
\partial _\eta {\tilde B}^+ &=& (- \gd k^2 +\gb k) {\tilde B}^+ 
                                     + \go k^2 {\tilde v}^+,
\label{tildeB+}
 \\
\partial _\eta {\tilde B}^- &=& (- \gd k^2 -\gb k) {\tilde B}^- 
                                     + \go k^2 {\tilde v}^-.
\label{tildeB-}
\end{eqnarray}
We multiply the first equation by ${\tilde B}^{+*}$ and the second
by ${\tilde B}^{-*}$ and take ensemble averages to get
\begin{eqnarray}
 \partial _\eta |B^+|^2 &=& 2 (- \gamma_D k^2 +\gamma_B k) |B^+|^2 +
     2 \gamma_\omega k^2 \langle {\tilde B}^{+*} {\tilde v}^+ \rangle ,
\label{eq:evo_bplus0}
 \\
 \partial _\eta |B^-|^2 &=& 2 (- \gamma_D k^2 -\gamma_B k) |B^-|^2 +
     2 \gamma_\omega k^2 \langle {\tilde B}^{-*} {\tilde v}^- \rangle .
\label{eq:evo_bminus0}
\end{eqnarray}
The magnetic field is zero initially and only the $\chi\omega$ term
is important on the right-hand side. Hence at early times, the 
solution to Eqs.~(\ref{tildeB+}) and (\ref{tildeB-}) is
\begin{equation}
{\tilde B}^{\pm}(\eta,{\bm k}) = \gamma_\omega k^2 
         \int_{\eta_0}^\eta d\eta' ~ {\tilde v}^{\pm} (\eta',{\bm k} ) 
\end{equation}
Note that $\gamma_\omega$ is, in principle, a function of time
since $\Delta\mu$ can vary due to the chiral anomaly relation. 
However, the $\chi\omega$ term cannot change the helicity
of the magnetic field as can be checked by taking the difference
of (\ref{eq:evo_bplus0}) and (\ref{eq:evo_bminus0}), and so
$\gamma_\omega$ can be assumed constant and taken out of the
integral. Then for the $\chi\omega$ term we get
\begin{equation}
\langle {\tilde B}^{\pm*}(\eta,{\bm k}) {\tilde v}^\pm (\eta,{\bm k}') \rangle
 = \gamma_\omega k^2 \int_{\eta_0}^\eta d\eta' ~ 
 \langle {\tilde v}^{\pm*} (\eta',{\bm k} ) 
                  {\tilde v}^\pm (\eta,{\bm k}' )\rangle
\label{Bvcorrelint}
\end{equation}
and we need the unequal time correlator for the velocity field.

We expect the fluid velocity at any $k$ mode to be correlated
on the eddy turnover time scale $2\pi/kv(\eta,k)$, where $v(\eta,k)$
is the fluid velocity, and to be uncorrelated on longer
time scales. This suggests
\begin{equation}
\langle {\tilde v}^{\pm*} (\eta', {\bm k} ) 
               {\tilde v}^\pm (\eta,{\bm k}' )\rangle
= \langle {\tilde v}^\pm (\eta,{\bm k} )^2 \rangle ~ 
      (2\pi)^3 \delta^{(3)} ({\bm k}-{\bm k}'),
\ {\rm for}\ |\eta - \eta'| < \frac{2\pi}{k v(\eta,k)}
\end{equation}
and 
\begin{equation}
\langle {\tilde v}^{\pm*} (\eta', {\bm k} ) 
               {\tilde v}^\pm (\eta,{\bm k}' )\rangle = 0, \ 
    {\rm for}\ |\eta-\eta'| > \frac{2\pi}{k v(\eta,k)}
\end{equation}
where $v(\eta,k)$ is the fluid velocity at length scale
$2\pi/k$ at time $\eta$. (We will relate $v(\eta,k)$ to
$v^\pm (\eta,k)$ in Eq.~(\ref{v+-nohel}) below.)

Inserting these unequal time correlators in Eq.~(\ref{Bvcorrelint}) gives
\begin{equation}
\langle {\tilde B}^{\pm*}(\eta,{\bm k}) 
      {\tilde v}^\pm (\eta,{\bm k}') \rangle
 = \gamma_\omega k^2  f(\eta,k) ~ |v^\pm|^2 ~
                     (2\pi)^3 \delta^{(3)} ({\bm k}-{\bm k}')\ .
\label{Bvcorrel}
\end{equation}
where $f(\eta ,k)=\eta-\eta_0$ for $\eta-\eta_0 \le 2\pi/(k v)$, and 
$f(\eta,k)=  2\pi/(kv)$ for $\eta-\eta_0 > 2\pi/(kv)$. 
An example of a 
smooth function with the same asymptotic properties is
\begin{equation}
f(\eta,k) = S \frac{2\pi}{kv}
 \tanh \left ( \frac{kv}{2\pi S} (\eta-\eta_0) \right )
\label{fadopted}
\end{equation}
where $S \sim 1$ is a fudge factor. 
We will adopt this choice for $f(\eta,k)$ in our numerical calculations. 

The form of the correlator (\ref{Bvcorrel}) is based on 
(\ref{Bvcorrelint}) which is valid at early times when the $\chi\omega$ 
effect dominates. The correlator will not be valid at later times when 
the $\chi\omega$ effect becomes sub-dominant, but then the form of the
correlator is also not important. Hence we can adopt the form
(\ref{Bvcorrel}) for all times.

So the MHD equations become
\begin{eqnarray}
 \partial _\eta |B^+|^2 &=& 2 (- \gamma_D k^2 +\gamma_B k) |B^+|^2 +
     2 \gamma_\omega^2 k^4 f(\eta,k) |v^+|^2
\label{eq:evo_bplus}
 \\
 \partial _\eta |B^-|^2 &=& 2 (- \gamma_D k^2 -\gamma_B k) |B^-|^2 +
     2 \gamma_\omega^2 k^4 f(\eta,k) |v^-|^2
\label{eq:evo_bminus}
\end{eqnarray}
These equations are linear in the quadratic variables $|B^\pm|^2$. 

From Eq.~(\ref{eq:mu-evo1}), the evolution of the ensemble averaged
comoving chemical potential is given by
\begin{equation}
 {d \Delta \mu  \over d\eta} = - c_\Delta \alpha
\int {k d k  \over 2 \pi^2} \partial _\eta 
                    \left [ |B^+|^2 - |B^-|^2 \right ]
       - \Gamma_F \Delta \mu ~ .
\label{eq:mu-evo}
\end{equation}

Eqs.~(\ref{eq:evo_bplus}), (\ref{eq:evo_bminus}) and (\ref{eq:mu-evo}) are
the dynamical equations that we will need to solve after specifying the
velocity modes, $v^\pm$.

\subsection{Fluid Velocity}
\label{subsec:fluidvelocity}

We consider the scenario of a strong first order phase transition at an 
early epoch, much earlier than the electroweak phase transition. As bubbles 
of the new phase grow and collide, the cosmological medium gets pushed and 
turbulence is generated. The radius of the bubbles at completion of the 
phase transition defines the length scale at which the fluid is being driven. 
Let us call this length scale $\zeta_b$.
In a non-expanding spacetime, on length scales smaller than $\zeta_b$
we expect a Kolmogorov distribution of velocities, and on larger 
scales, if we assume a random superposition of constant
velocity domains of size $\zeta_b$,
we expect a white noise spectrum. In a cosmological setting,
we also need to account for Hubble expansion. The largest scale relevant
for turbulent flows will be the size of the eddy that circulates
at least once per Hubble time. This length scale is $\zeta_H \sim v \eta$
where $v$ is the flow velocity driven by the bubbles.
Then the inertial scale is set by the smaller of $\zeta_b$ and $\zeta_H$ and 
will be denoted by $\zeta_i$.

The fluid velocity spectrum is given by a power law
\begin{equation}
  E_V(\eta, k) =  v_i^2(\eta) \left( {k \over k_i(\eta)}\right)^{n} ,
\label{eq:velocity_spectrum}
\end{equation}
where $k_i$ denotes the inertial wavenumber, $v_i$ is the fluid
velocity at this scale. On small length scales, $k>k_i$, the 
Kolmogorov spectrum gives $n=-2/3$, whereas we take a white noise spectrum, 
$n=3$, on large length scales, $k< k_i$. (The assumption
of white noise on large length scales is not crucial for the
main qualitative features in the evolution as these all occur
in the Kolmogorov part of the spectrum.)
From Eq.~(\ref{eq:velocity_spectrum})
we get the velocity at wavenumber $k$,
\begin{equation}
  v(\eta, k) =  v_i(\eta) \left( {k \over k_i(\eta)}\right)^{n/2}.
\label{eq:v_mode}
\end{equation}
In the radiation era, and on large length scales with $k < k_i (\eta )$,
the velocity field does not change 
({\it e.g.} Sec.~7.3.2 of \cite{Mukhanov:2005sc}). So
\begin{equation}
v(\eta, k) = v(\eta_0,k),\ \ k < k_i(\eta),
\label{eq:vk}
\end{equation}
where $\eta_0$ is the time of the phase transition.
Combining Eqs.~(\ref{eq:v_mode}) and (\ref{eq:vk}) with $n=3$, we obtain
\begin{equation}
v_i (\eta) =  v_i(\eta_0) \left( \frac{k_i(\eta )}{k_i(\eta_0)}\right)^{3/2}.
\label{eq:vi_mode}
\end{equation}
Now the inertial wavenumber, $k_i$, is defined by
\begin{equation}
k_i (\eta) =  \frac{2\pi}{v_i(\eta) \eta}.
\label{eq:ki}
\end{equation}
When inserted in Eq.~(\ref{eq:vi_mode}) we get
\begin{equation}
v_i (\eta) =  v_i (\eta_0) \left ( \frac{\eta_0}{\eta} \right )^{3/5}.  
\label{eq:vieta}
\end{equation}
Plugging back into (\ref{eq:ki}) gives \cite{Jedamzik:2010cy}
\begin{equation}
k_i (\eta) = k_i (\eta_0) \left ( \frac{\eta_0}{\eta} \right )^{2/5}.
\end{equation}
With these expressions the fluid velocity in Eq.~(\ref{eq:v_mode}) can
be written as
\begin{equation}
v(\eta, k) = v_i(\eta_0) \left ( \frac{\eta_0}{\eta} \right )^{(3-n)/5}
 \left ( \frac{k}{k_i(\eta_0)} \right )^{n/2}.
\label{eq:vfinal}
\end{equation}

We will assume that there is no kinetic helicity in the velocity
flow and from Eq.~(\ref{fluidkin}) write
\begin{equation}
|v^+(\eta,k)| = |v^-(\eta,k)| = \pi k^{-3/2}  v(\eta,k).
\label{v+-nohel}
\end{equation}

\begin{figure}
   \includegraphics[width=90mm]{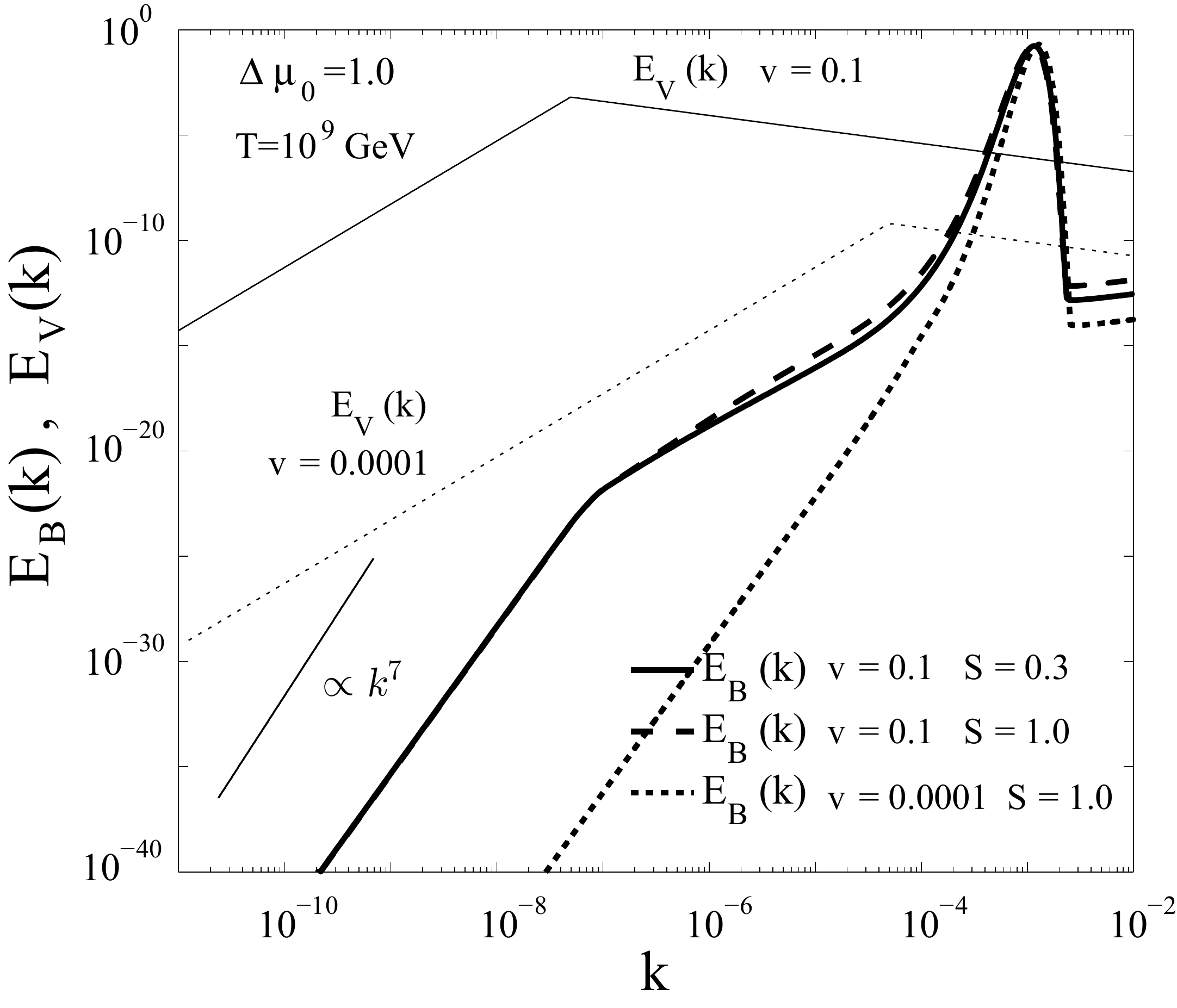}
\caption{Plots of the magnetic energy spectrum at $T=10^9$ GeV for
different values of the peak velocity and fudge factor $S$ (see 
Eq.~(\ref{fadopted})) exhibit very similar features.
}
\label{mt9comparison}
\end{figure}

\begin{figure}
 \begin{tabular}{cc}
 \begin{minipage}{0.5\hsize}
  \begin{center}
   \includegraphics[width=90mm]{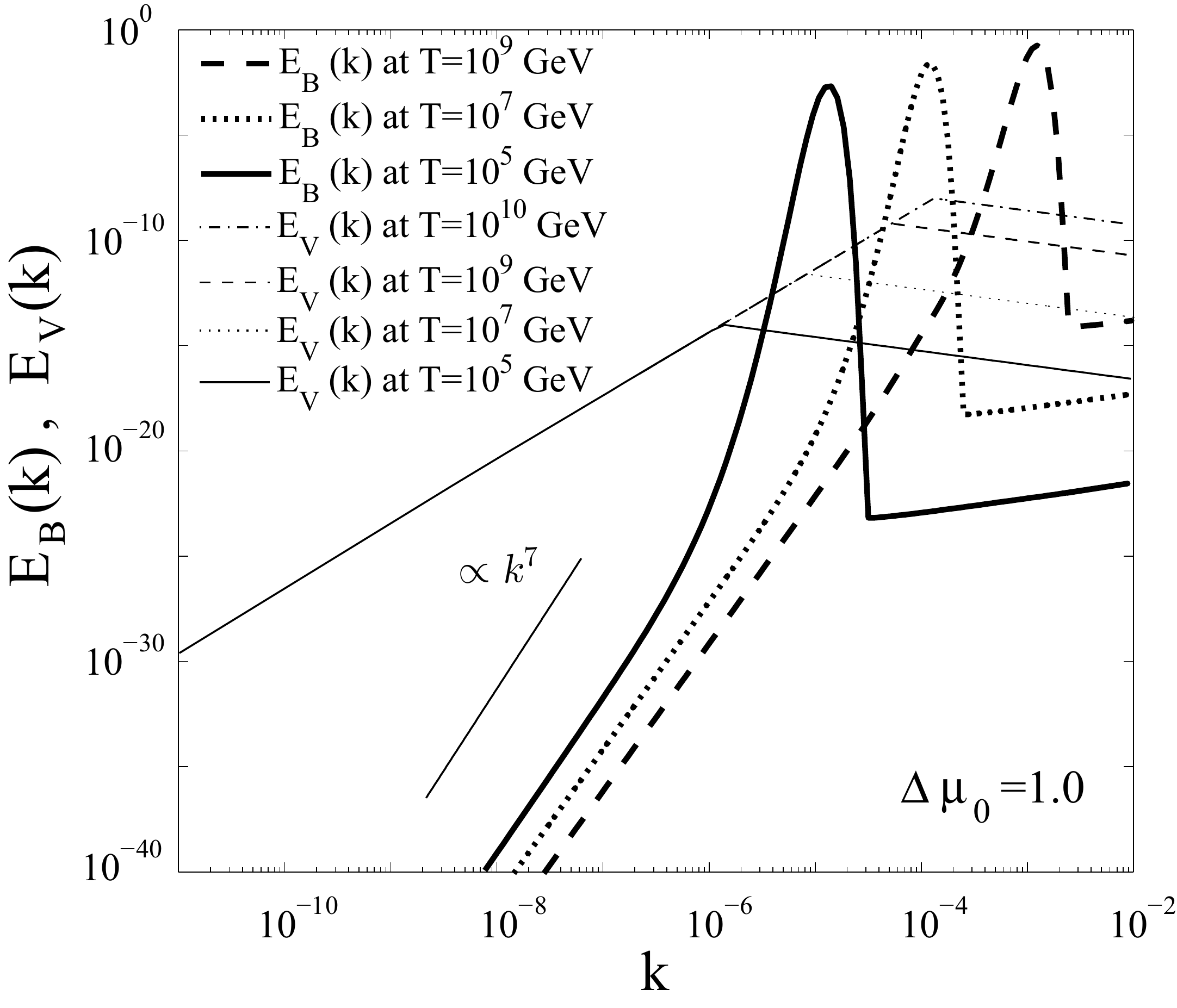}
  \end{center}
 \end{minipage}
 \begin{minipage}{0.5\hsize}
  \begin{center}
   \includegraphics[width=90mm]{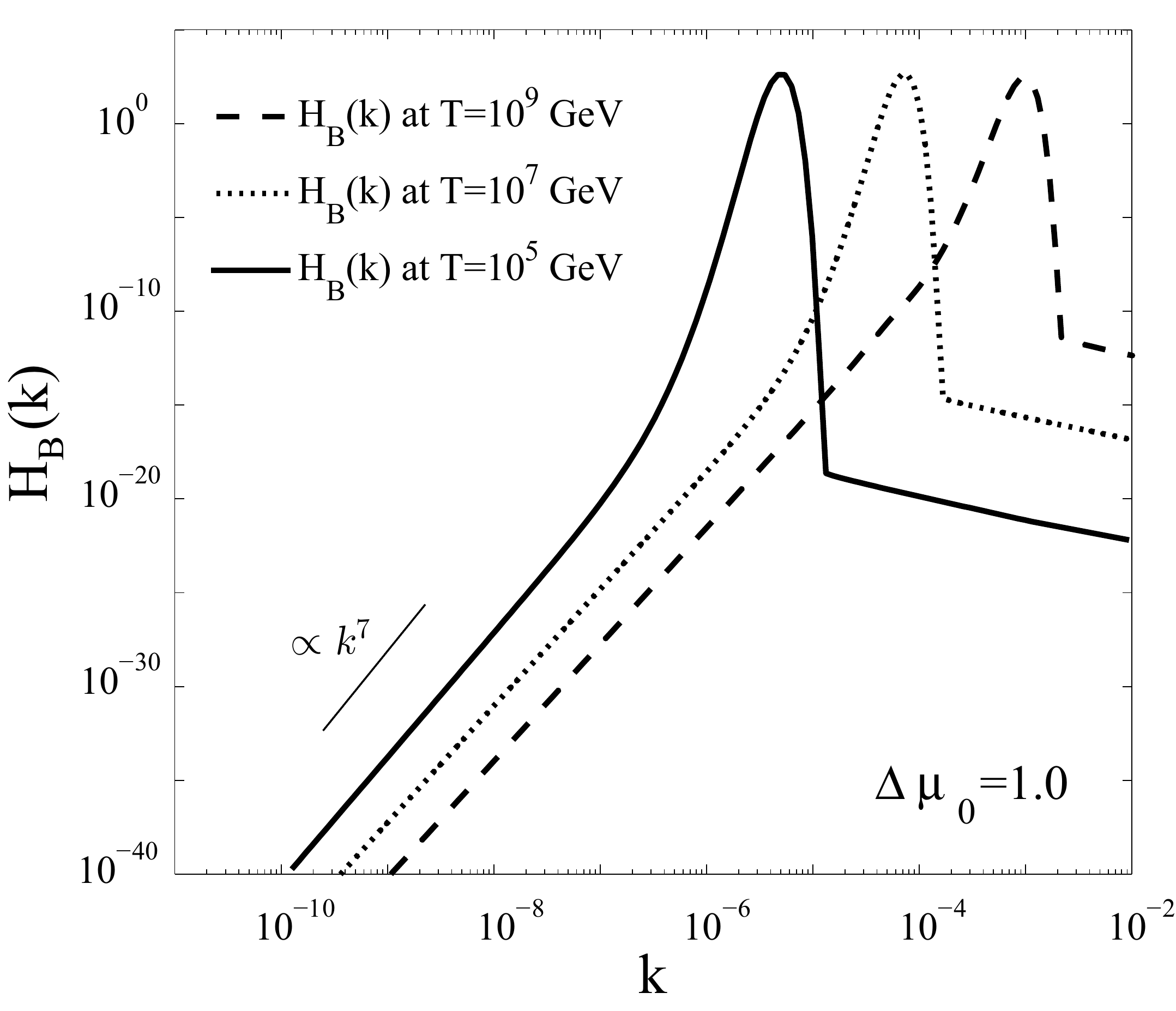}
  \end{center}
 \end{minipage}
\end{tabular}
\caption{Evolution when the initial velocity flow is non-helical with
$v_i(\eta_0)=10^{-4}$ and the initial chemical potential is 
$\Delta \mu_0 =1$.
The left panel shows the power spectra, $E_B$ and $E_V$, which
are normalized to the comoving radiation energy density, $g_* \pi^3/15$.
The thick lines are for $E_B$ and the thin lines are for $E_V$. 
The spectra are shown at the three different times, $T=10^9~{\rm GeV}$
(dashed), $T=10^{7}~{\rm GeV}$ (dotted) and $T=10^{5}~{\rm GeV}$
(solid). The thin dotted-dashed line is the initial velocity
power spectrum at $T=10^{10}~{\rm GeV}$ while the initial magnetic
field vanishes. Similarly, the right panel shows the evolution of
the magnetic helicity spectrum.
}
\label{fig:p_nohel_high}
\end{figure}

\begin{figure}
 \begin{tabular}{cc}
 \begin{minipage}{0.5\hsize}
  \begin{center}
   \includegraphics[width=90mm]{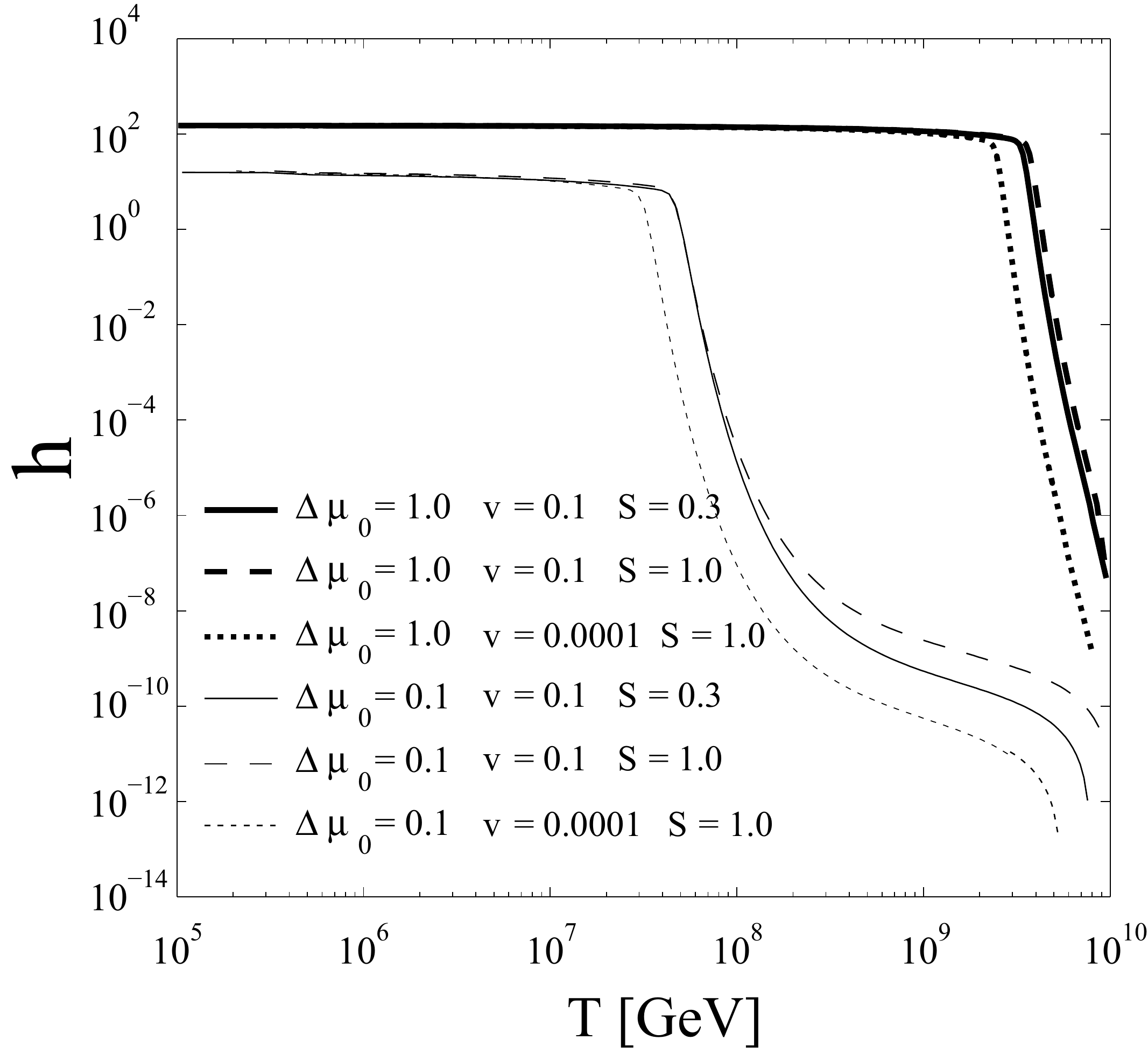}
  \end{center}
 \end{minipage}
 \begin{minipage}{0.5\hsize}
  \begin{center}
   \includegraphics[width=90mm]{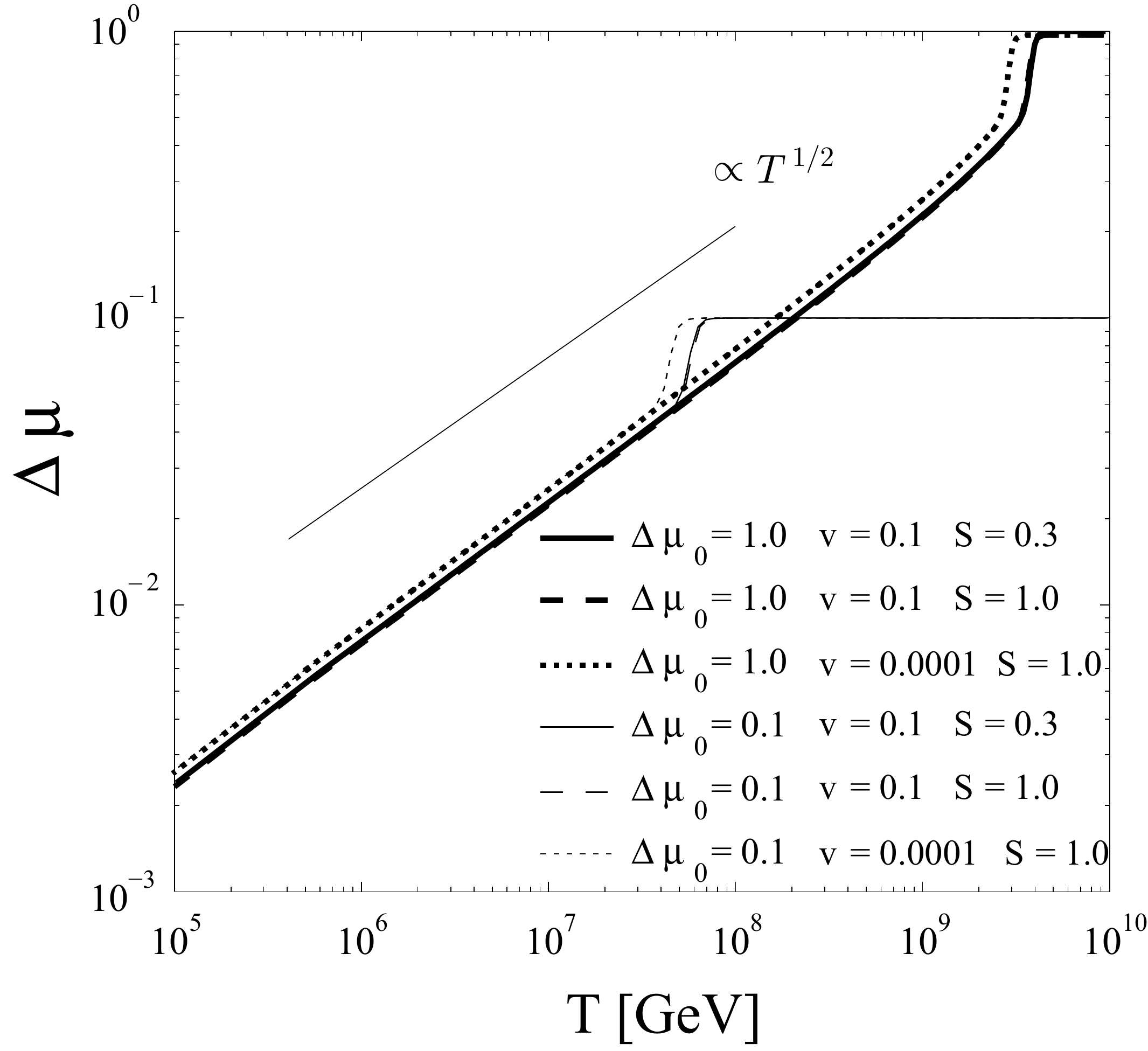}
  \end{center}
 \end{minipage}
\end{tabular}
  \caption{Evolution of the total helicity (left panel) and the chemical
potential (right panel), when the initial velocity flow is non-helical and
the initial magnetic field vanishes, for several different parameters.
}
\label{fig:mu_evo}
\end{figure}

\subsection{Numerical Evolution}
\label{sec:numerics}

We will now solve  Eqs.~(\ref{eq:evo_bplus}), (\ref{eq:evo_bminus})  and
(\ref{eq:mu-evo}) with the velocity spectra as given in 
Eqs.~(\ref{v+-nohel}). We will consider a few different values of 
$v_i(\eta_0)$ and take $\eta_0 = M_*/T_0$ 
where $T_0 =10^{10}~{\rm GeV}$ is the temperature at which the phase transition 
occurs and $M_*=6.6\times 10^{17}~{\rm GeV}$.
At the high temperatures we are considering, $e$ should be 
the Abelian (hypercharge) gauge coupling constant but, for numerical 
purposes, we 
take it to be given by $e=\sqrt{ 4 \pi \alpha}$ in $\go$ and $\gb$, 
where $\alpha =1/137$.
The comoving electrical conductivity is $\sigma =70$ as in 
Eq.~(\ref{sigmavalue}).
We consider several different values for the chemical potential at 
$\eta_0$ but focus on  $\Delta\mu_0=1.0=\Delta\mu_0^2$. 
The initial magnetic field is taken to vanish in all cases, as is the
flipping rate $\Gamma_F$ since this is small at temperatures above
$\sim 80~{\rm TeV}$ \cite{Campbell:1992jd}.
Note that the equations of motion are independent of the initial
epoch of turbulence but the velocity field in Eq.~(\ref{eq:vfinal})
explicitly contains $\eta_0$, and hence $T_0$.

We start by showing the magnetic energy spectrum at a fixed time
($T =10^9~ {\rm GeV}$) for several different values of the peak velocity
and parameter $S$ (defined in (\ref{fadopted})) in 
Fig.~\ref{mt9comparison}. The plots show that a sharp peak in
the spectrum develops and its position and shape are not very
sensitive to the input parameters. This can be understood from the evolution
equations (\ref{eq:evo_bplus}) and (\ref{eq:evo_bminus}). For $\gb >0$, we 
find that $B^-$ stays small and only
$B^+$ contributes to the magnetic energy density and helicity. At early
times, the velocity field acts as a source term for $B^+$ and the first
term on the right-hand side of Eq.~(\ref{eq:evo_bplus}) is negligible. 
However, with evolution, the first term becomes more important and the 
coefficient is negative for $k > \gb/\gd$ and positive for $k < \gb/\gd$.  
This change in sign implies a peak in the spectrum at 
$k \approx \gb/\gd \sim e^2 \Delta\mu/2\pi^2$, or on a comoving length scale
$4\pi^3/(e^2 \Delta\mu)$. The position of the peak at $k\approx \gb/\gd$
also agrees with our numerical results. We will give a more precise
analytical derivation of the peak location in Sec.~\ref{analytics}.

The left panel of Fig.~\ref{fig:p_nohel_high} represents the evolution of
$E_B(k)$ and $E_V(k)$, and the right panel shows the evolution of
$H_B(k)$, for $\Delta\mu_0=1$ and $v_i = 10^{-4}$. 
As is evident in Fig.~\ref{fig:p_nohel_high}, the 
peak position shifts toward large length scales with evolution. This can be 
understood by noting that the peak position $\propto \Delta\mu$, and 
$\Delta\mu$ is a decreasing function of time.

\begin{figure}
  \includegraphics[height=0.4\textwidth,angle=0]{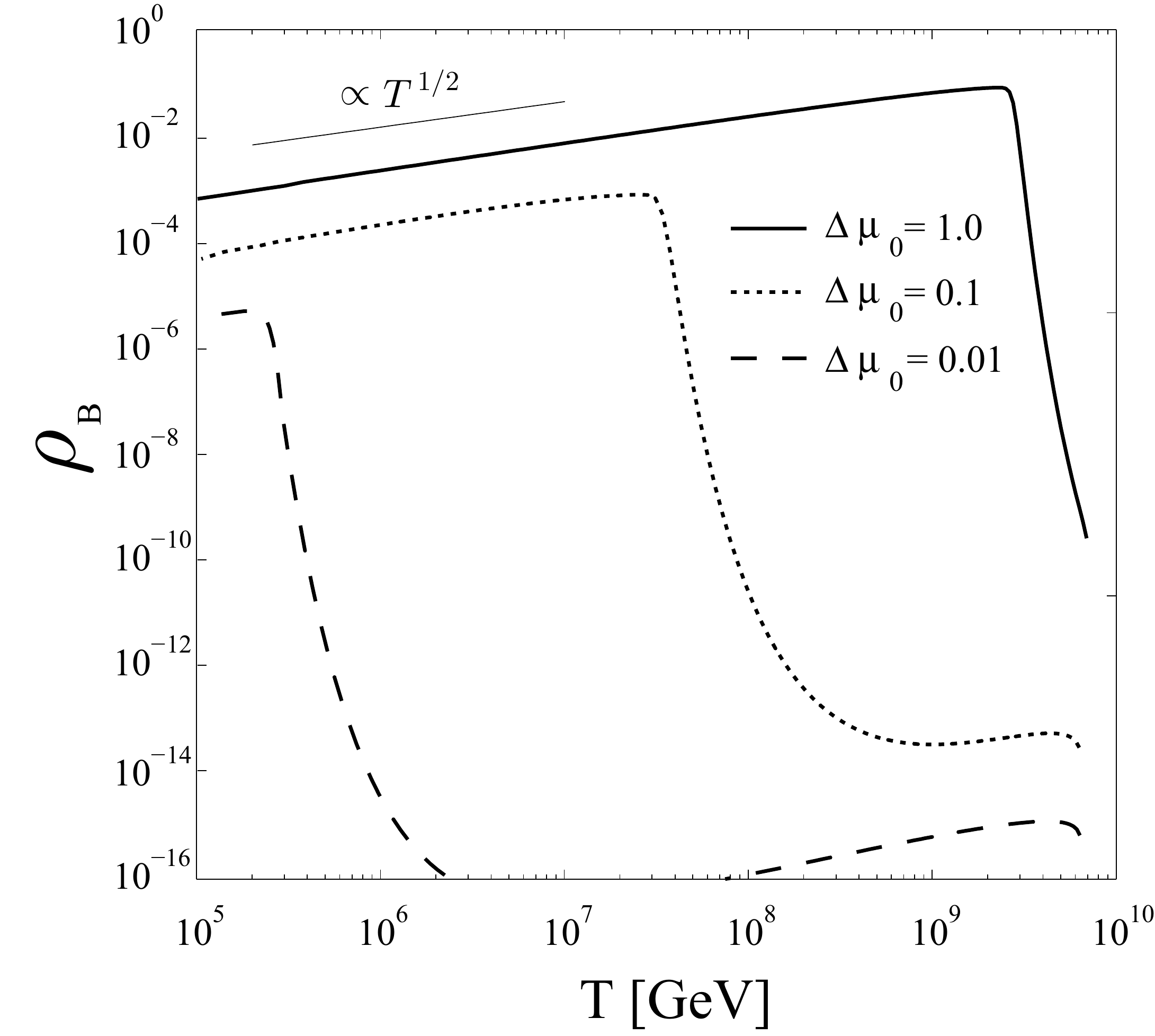}
\caption{
Evolution of magnetic energy density with temperature for
three different values of $\mu_0$ and with $v_i(\eta_0)=10^{-4}$.
}
\label{energyevo}
\end{figure}

The evolution of the chemical potential and magnetic helicity
are shown in the two panels in Fig.~\ref{fig:mu_evo}.  
The difference of chemical potential, $\Delta\mu$, decays and becomes
smaller, while the total helicity, $h$, evolves to its asymptotic 
value 
\begin{equation}
h_* = \Delta \mu_0/\alpha c_\Delta. 
\label{h*}
\end{equation}
The main contribution to the total magnetic helicity comes from the 
peak of the helicity spectrum. At this peak, the diffusion and 
$\chi B$ terms in Eq.~(\ref{eq:evo_bplus}) cancel each other. Further 
evolution occurs since the chemical potential is changing and this shifts
the peak position. 

The chemical potential difference $\Delta\mu$ has an ``attractor''
solution -- different initial values, $\Delta\mu_0$, all evolve to 
decay along a common trajectory with $T^{1/2}$ slope (see 
Fig.~\ref{fig:mu_evo}). The attractor behavior we observe is stronger 
than the ``tracking'' observed in \cite{Boyarsky:2011uy}  where the 
authors find constant helicity at late times in the case of a single 
magnetic field mode. Joyce and Shaposhnikov \cite{Joyce:1997uy} also 
observed the $T^{1/2}$ decay of $\Delta\mu$; in addition, we find
that the coefficient of the decay is independent of the initial
value of $\Delta\mu$. Eventually $\Delta\mu$ becomes very small and 
the magnetic helicity reaches its asymptotic values (\ref{h*}) as 
given by the chiral anomaly equation. We will obtain a better 
understanding of these features in the next section.

In Fig.~\ref{energyevo} we plot the magnetic energy
density as a function of temperature. During the attractor
evolution, this gives $\rho_B \propto T^{1/2}$.

Up to now, we have neglected the advection term in the magnetic field
evolution. This assumption is certainly valid in the early stages of 
the evolution when the magnetic field strength is very small and
the chemical potential is large.  At later times, however, we
might expect the advection term to become important.
To clarify this issue we compare the advection and $\chi B$ terms
at the peak position after we have analytically understood the
evolution in Sec.~\ref{analytics}.

Notice in Fig.~\ref{fig:p_nohel_high} that the magnetic energy density 
at the peak is larger than the kinetic energy density. One might 
expect backreaction of the magnetic fields on the fluid flow
to become important in this situation, and the magnetic and
kinetic spectra to asymptotically approach equipartition. 
However, in our case, except for a brief initial phase, the current 
is proportional to the magnetic field and so the Lorentz force,
${\bf j}\times {\bf B}$, vanishes. Thus there is no back-reaction
of the magnetic field on the fluid flow.

\subsection{Analytical Understanding}
\label{analytics}

In this section we will be able to understand most of the features
of the numerical evolution. Our analysis also applies to the 
scenarios considered in Refs.~\cite{Joyce:1997uy,Boyarsky:2011uy}
since the only difference is in the early stages of the evolution. 
In Ref.~\cite{Joyce:1997uy}, $\Delta\mu_0\ne 0$ and a non-zero
initial magnetic field are assumed. In Ref.~\cite{Boyarsky:2011uy}, 
$\Delta\mu_0=0$ and the initial helical seed field is injected during
baryogenesis. In the present work $\Delta\mu_0\ne 0$ and a non-zero 
seed field is generated due to turbulence and the $\chi\omega$ effect.

The equations to be evolved are (\ref{eq:evo_bplus}),
(\ref{eq:evo_bminus}) and (\ref{eq:mu-evo}). Initially,
$B^\pm=0$ and so only the $\chi\omega$ terms are important.
Then
\begin{equation}
|B^\pm (\eta,k)|^2 \approx 2 k^4 
  \int_{\eta_0}^\eta d\eta ~ \gamma_\omega^2 
               f(\eta , k)~ |v^\pm (\eta, k)|^2 
\label{chiomB}
\end{equation}
Note that $\gamma_\omega$ has not been pulled out of the time
integration because it is proportional to $\Delta\mu$ and
this must be evolved according to (\ref{eq:mu-evo}). However,
at these early stages, $B^+ \approx B^-$ and the right-hand
side of (\ref{eq:mu-evo}) vanishes, and so $\Delta\mu$ and
$\gamma_\omega$ are constant. This is also seen in 
Fig.~\ref{fig:mu_evo} where the curves of $\Delta\mu$ are 
flat at the highest temperatures. 

The $\chi\omega$ dominated evolution continues until 
the $\chi B$ term starts to become important. The transition
occurs at time $\eta_1$ such that
\begin{equation}
\gamma_B k \int_{\eta_0}^{\eta_1} d\eta ~ 
                     f(\eta,k) |v^\pm (\eta, k)|^2
              ~ \approx ~ f(\eta_1,k) |v^\pm (\eta_1,k)|^2
\label{eq:eta1}
\end{equation}
where $v^\pm$ are given in Eq.~(\ref{v+-nohel}).
For $\eta > \eta_1$, we can ignore the $\chi\omega$ term. 

In the Kolmogorov part of the spectrum, Eq.~(\ref{eq:vfinal})
gives $v^\pm \propto \eta^{-11/15}$ and, assuming
$2\pi \gamma_B < 1$, the integral on the 
left-hand side of Eq.~(\ref{eq:eta1}) is dominated by the 
contribution from $\eta \sim 2\pi /k$. 
A simple estimate then gives
\begin{equation}
\eta_1 \approx \frac{1}{(2\pi\gamma_B)^{15/22}} \frac{2\pi}{k}
\end{equation} 
and this introduces an extra $k$ dependence in $B^\pm (\eta_1,k)$. 
However, we find $B^\pm (\eta_1(k),k) \propto k^{-1/10}$ in 
the Kolmogorov part of the spectrum, and this weak dependence 
will be ignored in what follows. Also, as in 
Fig.~\ref{fig:p_nohel_high}, the evolution will lead to a 
sharp peak in the spectrum, which is the most interesting 
and dominant feature in the spectrum. In discussing this peak, 
we can treat $\eta_1$ as being independent of $k$, since
the Kolmogorov spectrum is relatively flat over the width
of the peak. So we will simply set $\eta_1=0$ from now on. 
This simplification will not affect the gross features of the 
evolution.

Ignoring the $\chi\omega$ term, the equation for $B^\pm$ can be 
written as
\begin{equation}
\partial_\eta |B^\pm| = -\gamma_D k (k\mp k_p) |B^\pm | 
\label{B+nochiomega}
\end{equation}
with 
\begin{equation}
k_p \equiv \frac{\gamma_B}{\gamma_D} =
\frac{e^2 \Delta\mu}{2\pi^2}\ .
\label{kpDeltamu}
\end{equation}
Eq.~(\ref{B+nochiomega}) can be solved and the solution written
as the product of an initial amplitude, an exponential amplification
factor common to all modes, and a Gaussian spectrum,
\begin{equation}
|B^\pm (\eta,k)| = |B^\pm_0| \exp (\gamma_D K_p^2 \eta )
\exp \left [ -\gamma_D \left ( k \mp K_p \right )^2 \eta \right ]
\end{equation}
where the peak of the Gaussian for $B^+$ is at
\begin{equation}
K_p \equiv \frac{1}{2\eta}\int_{0}^{\eta} d\eta ~ k_p
\label{Kpdefn}
\end{equation}
The restriction $k > 0$ implies that
$B^-$ is given by the tail of a Gaussian which is peaked at 
$k=-K_p$. Therefore $B^-$ can be ignored except for $\eta \approx 0$. 

Thus the spectrum of $B^+$ is a Gaussian of width
\begin{equation}
\Delta k \approx \frac{1}{\sqrt{\gamma_D\eta}}
\end{equation}
which is consistent with the width of the peaks in 
Fig.~\ref{fig:p_nohel_high}. It is also interesting to note that
the peak gets narrower with time.

Now we can estimate the magnetic helicity
\begin{eqnarray}
h &=& \int \frac{dk}{2\pi^2} k  (|B^+|^2 - |B^-|^2) \\
 &\approx&  \frac{A}{\sqrt{\eta}}  K_p  e^{2\gamma_D K_p^2\eta} 
\label{helestimate}
\end{eqnarray}
where $A =|B^+_0|^2/\sqrt{2\pi^2 \gamma_D}$.
The magnetic energy density is obtained from Eq.~(\ref{EB(k)})
\begin{eqnarray}
\frac{1}{2} \langle |{\bm B}(\eta, {\bm x})|^2 \rangle
&=& \int {dk \over 4 \pi^2} k^2 [ \langle |B^+ (\eta, k)|^2 \rangle 
                    +\langle |B^- (\eta, k)|^2 \rangle] \nonumber\\
 &\approx&  \frac{A}{2\sqrt{\eta}}  K_p^2  e^{2\gamma_D K_p^2\eta} 
\label{energyestimate}
\end{eqnarray}

From the chiral anomaly equation, we also have
\begin{eqnarray}
\Delta\mu &=& \Delta\mu_0 - c_\Delta \alpha h\\
          &\approx & \Delta\mu_0 - 
 \frac{c_\Delta \alpha A}{\sqrt{\eta}}  K_p  e^{2\gamma_D K_p^2\eta} 
\label{Dmuestimate}
\end{eqnarray}

As discussed above, during the $\chi\omega$ dominated phase, the
helicity stays near zero and $\Delta\mu \approx \Delta\mu_0$. 
Then $k_p\approx e^2 \Delta\mu_0/2\pi^2$ 
and $K_p \approx k_p/2 \approx 
{\rm constant}$.
In a time $\approx [\gamma_D (\Delta\mu_0)^2]^{-1}$,
the exponential in (\ref{Dmuestimate}) starts to become important.
Then the magnetic helicity increases exponentially fast and 
$\Delta\mu$ rapidly decreases. 

The exponential growth of the helicity terminates when it gets close
to the asymptotic value $h_*$, given by Eq.~(\ref{h*}).  The subsequent
evolution can be investigated by setting $h\approx h_*$ in
Eq.~(\ref{helestimate}).  In order for $h$ to saturate at $h_*$, the exponent
in Eq.~(\ref{helestimate}) has to become nearly constant, $K_p^2 \eta\approx
{\rm const}$, or
\begin{equation}
K_p \propto \eta^{-1/2}.
\end{equation}
Then Eqs.~(\ref{Kpdefn}) and (\ref{kpDeltamu}) imply
\begin{equation}
k_p\approx K_p
\end{equation}
and
\begin{equation}
\Delta\mu \propto \eta^{-1/2} \propto T^{1/2}
\end{equation}
which is the attractor solution seen numerically.

To make these estimates more quantitative, we set 
\begin{equation}
K_p \approx k_p = \frac{e^2 \Delta\mu}{2\pi^2} \approx C T^{1/2},
\label{KCT}
\end{equation}
where $C$ is a constant (or, more exactly, a slowly varying function) to be
determined.
Substituting this into (\ref{helestimate}) and using $T=M_*/\eta$, we obtain
\begin{equation}
C \approx (2 \gamma_D M_*)^{-1/2} ({\rm ln} X)^{1/2},
\label{Ceq}
\end{equation}
where
\begin{equation}
X = (h_*/A K_p)\sqrt{\eta} \propto \eta.
\end{equation}
This shows that $C$ has a weak, logarithmic dependence on $\eta$.  

Comparing Eqs.~(\ref{helestimate}) and (\ref{energyestimate}) and using 
(\ref{KCT}), we can estimate the magnetic energy density in the attractor 
regime,
\begin{equation}
\frac{1}{2} \langle |{\bm B}(\eta,{\bm x})|^2 \rangle \approx 
\frac{1}{2} h_* K_p
\approx \frac{C \Delta\mu_0}{2\alpha c_\Delta} T^{1/2}.
\label{Battractor}
\end{equation}
Note that, apart from the logarithmic factor, the coefficient $C$ in
(\ref{KCT}) is a constant independent of $\Delta\mu_0$ and of the turbulent
velocity spectrum $v^\pm (\eta, k)$.  Thus Eq.~(\ref{Battractor}) shows that
the attractor magnetic energy density is (roughly) proportional to the initial
chiral disbalance $\Delta\mu_0$ and is largely independent of $v^\pm (\eta,
k)$.  The characteristic length scale of the magnetic field,
\begin{equation}
K_p^{-1} \approx C^{-1}T^{-1/2},
\label{Kp-1}
\end{equation}
is not sensitive to any of the input parameters.

To get numerical estimates, we evaluated $X$ and $C$ at the lower end of 
our temperature range, $T=10^5$~GeV. This is done by solving the 
transcendental equation (\ref{Ceq}), after substituting expressions
for $A$ (below (\ref{helestimate})), $B^+_0$ (from (\ref{chiomB})), 
$\eta_1$ (from (\ref{eq:eta1})), and other relevant quantities. 
For $\Delta\mu_0=1.0$, this gives 
$X\approx 3.8\times 10^{16}$, $C\approx 4.5\times 10^{-8}$, in reasonable 
agreement with our numerical results.  
We have verified that for the range of $T$ and $\Delta\mu_0$ that we
considered here $C$ does not vary from this value by more than $0.8 \%$.

We now compare the advection term to the $\chi B$ term at the peak
position. The estimate in Eq.~(\ref{eq:terms}) shows that the advection term
is sub-dominant if
\begin{equation}
v_x \ll \gamma_B
\end{equation}
where $v_x$ denotes the velocity in physical space. 
We use Eq.~(\ref{fluidkin}) to go to momentum space
\begin{equation}
v_x \sim  v(\eta,k)
\end{equation}
and $v(\eta,k)$ is given in Eq.~(\ref{eq:vfinal}). Now we set
$k=K_p = CT^{1/2}$, $\eta =M_*/T$ in (\ref{eq:vfinal}) to get
\begin{equation}
v(\eta,K_p) = v_{i0} 
       \left ( \frac{CM_* v_{i0}}{2\pi T_0^{1/2}} \right )^{n/2}
       \left ( \frac{T}{T_0} \right )^{(3-n)/5 + n/4}
\end{equation}
where $v_{i0}=v_i(\eta_0)$ and $n=-2/3$. Inserting numbers in 
GeV units
$v_{i0}=10^{-4}$, $C=4.5\times 10^{-8}$, $M_*=6.6\times 10^{17}$,
$T_0 =10^{10}$ we get
\begin{equation}
v(\eta,K_p) = 6\times 10^{-5} \left ( \frac{T}{T_0} \right )^{17/30}
\end{equation}
We also have
\begin{equation}
\gamma_B = \frac{K_p}{\sigma} = 
\frac{C T_0^{1/2}}{\sigma} \left ( \frac{T}{T_0} \right )^{1/2}
= 6\times 10^{-5} \left ( \frac{T}{T_0} \right )^{1/2}
\end{equation}
The temperature dependence of $v$ and $\gamma_B$ are very similar
and so whatever relation holds at the initial time, also holds 
for all (relevant) later times. For a range of parameters
we have used in the numerical analysis, $v \lesssim \gamma_B$, 
and so the advection term is only marginally important. (From Eq.~(\ref{Ceq}) 
we see that $C$ is only sensitive to the parameter $\gamma_D = 1/\sigma$.)  
A smaller value of $v_{i0}$, or a larger value of $\gamma_B$ due to
several chiral particle species contributing to the $\chi B$ effect, or
a higher initial temperature $T_0$, can further ensure that the
advection term is unimportant.  We also note that the advection term
becomes negligible when turbulence is eventually dissipated on the relevant
scales. However, if the condition $v \ll \gamma_B$ is not satisfied, the
advection term will be important and a full numerical analysis will be 
necessary.

To summarize our understanding, at very early times, only
the $\chi\omega$ effect is important and this generates magnetic 
fields from the assumed turbulence. As these magnetic fields 
get stronger, the dissipation and $\chi B$ effects become
more important than the $\chi\omega$ effect. These two terms
take the initial spectrum produced by the $\chi\omega$ effect, 
and introduce a Gaussian peak in the spectrum at $k=K_p$
whose amplitude grows exponentially
and with width that decreases as $1/\sqrt{\eta}$. 
At the same time the magnetic helicity $h$ increases exponentially, the
magnetic field becomes maximally helical, and
$\Delta\mu$ rapidly decreases.  The rapid growth of $h$ terminates as it
approaches the asymptotic value $h_*$, and the system enters the attractor
regime, in which
$K_p \approx k_p \propto T^{1/2}$ and $\Delta\mu \propto T^{1/2}$. 
This means that the peak of the spectrum continues to evolve
to larger length scales. In the attractor evolution, the
magnetic helicity stays approximately constant 
while the magnetic energy density falls off as $T^{1/2}$.

\section{Conclusions}
\label{conclusions}

Maximal parity violation in the standard electroweak model suggests
that there may have been an asymmetry in the distribution of left- and 
right-handed chiral fermions in the early universe, so that the
cosmic medium was a ``chiral plasma''. Such a chiral asymmetry 
has implications for the generation of magnetic fields via the chiral-magnetic
and chiral-vorticity effects \cite{Vilenkin:1979ui,Vilenkin:1980fu}.
The latter effect leads to the production of magnetic fields during a 
turbulent phase in the chiral plasma, while the chiral-magnetic effect 
leads to amplification of the magnetic 
field \cite{Joyce:1997uy,Boyarsky:2011uy}.
These effects are in addition to the generation of magnetic fields 
during phase transitions \cite{Vachaspati:1991nm,Vachaspati:1994xc}
and during sphaleron processes responsible for baryogenesis 
\cite{Vachaspati:1994xc,Cornwall:1997ms,Vachaspati:2001nb,Copi:2008he,
Chu:2011tx,Vachaspati:2008pi}.

In this paper we have focused on the consequences of the $\chi\omega$ and
$\chi B$ effects. We have investigated the time-dependent solution
numerically in an expanding universe but with two assumptions. First we have 
considered negligible backreaction of the magnetic field on the
fluid velocity, and second we have assumed that the
advection term is unimportant. These assumptions simplify the analysis
because they linearize the MHD equation. However, the analysis is still
highly non-trivial because the chiral anomaly equation connects the
evolution of the chemical potentials to the total magnetic helicity
which is a sum over all magnetic modes.

Neglect of back-reaction is justified by the fact that maximally helical magnetic fields generated by our mechanism are nearly force-free.  We have also verified
that it is possible to choose parameters
such that the advection term is unimportant for the evolution of
the magnetic field at the peak of the spectrum. We leave the 
inclusion of the advection term, and perhaps a full-blown 
magneto-hydrodynamic analysis, for future work.

Our numerical solutions show that a magnetic field is generated due to 
the $\chi\omega$ effect in a turbulent chiral plasma. As the field gets
stronger, the $\chi B$ effect becomes important and leads to rapid 
amplification of the field on a preferred length scale given by 
$l_B \sim (K_p T)^{-1} \sim 4\pi^3/(e^2\Delta\mu)$.\footnote{In this section we use the physical, rather than `comoving' value for the chemical potential difference $\Delta\mu$.} The
$\chi\omega$ effect becomes insignificant at this stage.  The amplification
period ends when dissipation (magnetic diffusion) becomes important, and the
evolution approaches an attractor regime in which the chemical potential
difference decreases as $T^{1/2} \propto t^{-1/4}$, 
with a corresponding shift of power in the magnetic 
field to larger length scales.  The magnetic helicity density $h$ remains
nearly constant in this regime, and the magnetic energy density $\rho_B$
decreases as $T^{1/2}$.  The attractor solution for $\rho_B(t)$ is proportional
to the initial chiral disbalance $\Delta\mu_0/T_0$, where $T_0$ is the temperature at the phase transition, and has only a weak logarithmic
dependence on the magnitude of velocities and on the spectrum of turbulent
motion of the plasma.  The coherence length scale of the field is largely
independent of both $\Delta\mu_0/T_0$ and the turbulence spectrum.

The attractor regime ends at  $T_F \sim 80~{\rm TeV}$, when helicity flipping becomes important.  The magnetic energy density at this point is $\rho_B \sim  10^{-3} (\Delta\mu_0/T_0) \, \rho_r$ 
where  $\rho_r$ is the energy density in radiation, and the coherence length is $l_B \sim 
10^{5} T_F^{-1}$, which corresponds to the present comoving scale of $10^{4}$cm.
At $T < T_F$, chirality flipping will reduce $\Delta\mu/T$ and 
shift the peak to yet larger length scales as in Ref.~\cite{Boyarsky:2011uy}.
When the temperature drops to $\sim 0.1~{\rm MeV}$, electron-positron pairs annihilate, resulting in a sudden decrease of the electrical 
conductivity, $\sigma$, by a factor $\sim 10^{-9}$.  Analysis similar to that in Sec.~\ref{analytics} suggests
that a sudden decrease in $\sigma$ at a time 
when $\Delta\mu=0$ will decrease $K_p$ and thus increase the coherence
scale. In addition, the width of the Gaussian peak
in the spectrum will also increase.  Further growth of the coherence length may be caused by the inverse cascade in the decaying MHD turbulence (see, e.g., \cite{Banerjee,Campanelli} and references therein).  All these effects combined could shift the present coherence length into the range of astrophysical interest.  We hope to discuss the evolution of the spectrum through the various cosmological epochs separately.

In addition to the time-dependent analysis we have solved the MHD
equations for a steady state solution. This analysis points to the 
estimate in Eq.~(\ref{k0soln}) for the initial field generated by 
the cosmic turbulence.

\acknowledgments
We thank Avi Loeb and Oleg Ruchayskiy for useful comments. 
TV is grateful to Tufts Institute of Cosmology and to the 
Institute for Advanced Study, and AV thanks ASU for hospitality. 
TH thanks the ASU Advanced Computing Center for computing support.
This work was supported by DOE at ASU, and NSF at Tufts University.

\appendix

\section{Steady state solutions}
\label{app:steady}

To obtain the steady state solutions, we set $\partial_t {\bm B} =0$
in Eq.~(\ref{mhdeq}) and assume that the velocity field is stationary, $\partial_t {\bm v} =0$.

\subsection{Negligible advection}

We first consider the case when the advection term, 
${\bm \nabla}\times ({\bm v}\times {\bm B})$, is small compared to 
the other terms. Then Eq.~(\ref{mhdeq}) reduces to 
\begin{equation}
\gd \nabla^2 {\bm B} + \go {\bm \nabla} \times {\bm \omega} + 
\gb {\bm \nabla}\times {\bm B}=0,
\end{equation}
or
\begin{equation}
{\bm \nabla}\times [ -\gd {\bm \nabla}\times {\bm B} + \go {\bm \omega} + 
         \gb {\bm B} ]=0,
\end{equation}
Then
\begin{equation}
-\gd {\bm \nabla}\times {\bm B} + \go {\bm \omega} + \gb {\bm B} = {\bm \nabla}\phi,
\end{equation}
where $\phi$ is any scalar function. However, the divergence of the
left-hand side vanishes and hence we find $\phi =0$. So the equation
reduces to
\begin{equation}
-\gd {\bm \nabla}\times {\bm B} + \go {\bm \omega} + \gb {\bm B} =0.
\label{mhdeq2}
\end{equation}

In Fourier space 
\begin{equation}
{\bm B} = \int {d^3 k \over (2 \pi)^3} ~ {\bm B}_k e^{i{\bm k}\cdot {\bm x}},
\end{equation}
\begin{equation}
{\bm v} = \int {d^3 k \over (2\pi)^3} ~ {\bm v}_k e^{i{\bm k}\cdot {\bm x}}.
\end{equation}
Inserting in Eq.~(\ref{mhdeq2}) gives
\begin{equation}
-i \gd {\bm k}\times {\bm B}_k + i \go {\bm k}\times {\bm v}_k + \gb {\bm B}_k =0,
\label{eqinkspace}
\end{equation}
The dot product of this equation with ${\bm k}$ gives 
\begin{equation}
{\bm k}\cdot {\bm B}_k =0,
\end{equation}
and a dot product with ${\bm v}_k$ gives
\begin{equation}
(-i \gd {\bm v}_k \times {\bm k} + \gb {\bm v}_k)\cdot {\bm B}_k =0.
\end{equation}
Therefore
\begin{equation}
{\bm B}_k = F({\bm k}) {\bm k} \times
          (-i \gd {\bm v}_k \times {\bm k} + \gb {\bm v}_k).
\end{equation}
To determine the scalar function $F({\bm k})$, insert into
Eq.~(\ref{eqinkspace}) and obtain
\begin{equation}
F = \frac{-i \go}{\gb ^2- \gd ^2k^2}.
\end{equation}
Therefore
\begin{equation}
{\bm B}_k = \frac{-i \go}{\gb ^2-\gd ^2k^2} {\bm k} \times
          (-i\gd {\bm v}_k \times {\bm k} + \gb {\bm v}_k).
\label{advsoln}
\end{equation}

If we also assume that the fluid is incompressible, we get 
\begin{equation}
{\bm B}_k = \frac{-i \go}{\gb^2-\gd^2k^2}[\gb {\bm k}\times{\bm v}_k
                                    -i\gd k^2 {\bm v}_k ] \ , \ \ 
{\rm if}\ {\bm k}\cdot {\bm v}_k=0\ .
\end{equation}

We now return to our approximation in this section, namely to
ignore the advection term. As discussed at the end of the previous
section, this is justified for small fluid flow velocities and small 
length scales. Our solution shows that there is another situation 
when it is fair to ignore the advection term. Suppose the spectrum of
velocities is dominated by a single mode. Then the advection term
in Fourier space is ${\bm k} \times ({\bm v}_k \times {\bm B}_k)$ 
and this vanishes when we use the solution in Eq.~(\ref{advsoln}).

The solution (\ref{advsoln}) simplifies in the large wavelength 
limit, ${\bm k} \to 0$. Then
\begin{equation}
{\bm B}_k = -i \frac{\go}{\gb} {\bm k}\times {\bm v}_k,
\end{equation}
or
\begin{equation}
{\bm B} = - \frac{\go}{\gb} {\bm \omega} 
        = - \frac{\mu_L+\mu_R}{2e}{\bm \omega}.
\label{k0soln}
\end{equation}
This can also be seen directly from Eq.~(\ref{mhdeq}) because
we have discarded the advection term and the dispersion term
becomes negligible on large length scales. So the steady state
solution is given by
\begin{equation}
{\bm J}_{\chi\omega} + {\bm J}_{\chi B} =0,
\end{equation}
which leads to Eq.~(\ref{k0soln}).

To get an estimate for the magnetic field strength 
with coherence scale $L$, we assume
\begin{equation}
|{\bm \omega}_L| \sim \frac{v_L}{L}.
\end{equation}
This gives
\begin{equation}
B \sim - \frac{\mu_L + \mu_R}{2e L} v_L.
\end{equation}

In a time-dependent situation with ${\bf B}=0$ initially, we can expect a field 
of strength (\ref{k0soln}) to develop on a time scale
\begin{equation}
\tau_L\sim L/\gamma_B.
\label{tauL}
\end{equation}
This estimate follows by taking the ratio of the $\chi\omega$ term in
the evolution equation, $\sim \gamma_\omega \omega/L$, and the asymptotic 
value of the magnetic field $\sim (\gamma_\omega/\gamma_B) \omega$.
In a turbulent flow, individual eddies are not expected to persist 
much longer than a single revolution time, $\sim \omega_L^{-1}$. This 
is longer than $\tau_L$, provided that 
\begin{equation}
v_L < \gamma_B.
\end{equation}
Note that in this regime, the advection term in (\ref{mhdeq}) is small compared
to the $\chi B$ term, so the neglect of advection is justified.

If $v_L$ is given by the Kolmogorov spectrum 
\begin{equation}
v_L \propto v_t \left ( \frac{L}{t} \right )^{1/3},
\end{equation}
where $v_t$ is the velocity on a cosmological length scale, 
and assuming that $v_t \sim 1$ and $(\mu_L + \mu_R)\sim T$, we have
\begin{equation}
B \sim - \frac{T}{2e L^{2/3} t^{1/3}}.
\end{equation}
Inserting numbers suitable for the electroweak scale, we obtain
the magnetic field on the length scale $L$,
\begin{equation}
B_L \sim \frac{10^{18}}{(L T_{\rm EW})^{2/3}} ~{\rm G} ,
\label{B_L}
\end{equation}
where $T_{\rm EW} \sim 100~{\rm GeV}$.

\subsection{Negligible diffusion}
\label{sec:advecdom}

We now find the steady-state solution with the advection term
but without the diffusion term.
Now the steady state equation is
\begin{equation}
{\bm \nabla}\times ({\bm v}\times {\bm B})
  + \go {\bm \nabla}\times {\bm \omega} + 
\gb {\bm \nabla}\times {\bm B} = 0,
\label{eq:advecsteady}
\end{equation}
which leads to
\begin{equation}
{\bm v}\times {\bm B} + \go {\bm \omega} + \gb {\bm B} = {\bm \nabla}\phi,
\end{equation}
where $\phi$ can be any function, including a (scalar) function of the 
flow velocity e.g. ${\bm v}^2$. Unlike the case of diffusion domination,
the divergence of the left-hand side does not vanish and we cannot argue 
away $\phi$.

By taking scalar and vector products of ${\bm v}$ with 
Eq.~(\ref{eq:advecsteady}) we find the solution
\begin{equation}
{\bm B} = - \frac{\go}{\gb(\gb^2+v^2)} \left [ 
  ({\bm v}\cdot {\bm \omega}'){\bm v}-\gb {\bm v}\times{\bm \omega}'
      + \gb^2 {\bm \omega}' \right ] ,
\label{Bsolnadv}
\end{equation}
where
\begin{equation}
{\bm \omega}' = {\bm \omega}-\frac{1}{\go} {\bm \nabla}\phi.
\label{omega'}
\end{equation}

To determine $\phi$ we impose ${\bm \nabla}\cdot {\bm B} =0$. 
This gives the equation for $\phi$,
\begin{equation}
\partial_i \left [ 
\frac{( \delta^{ij} + u^i u^j )}{1+{\bm u}^2} \partial_j \Phi 
           \right ] - 
  \left [ {\bm \nabla}\times \left ( \frac{\bm u}{1+{\bm u}^2} \right ) 
  \right ] \cdot {\bm \nabla}\Phi = S,
\label{Phieq}
\end{equation}
where ${\bm u}={\bm v}/\gb$, $\Phi = \phi /\gb$ and
\begin{equation}
S = \go ~ {\bm \nabla}\cdot \left [ 
 \frac{ ({\bm u}\cdot {\bm \Omega} ) {\bm u} - 
          {\bm u}\times {\bm \Omega} + {\bm \Omega} }
          {1+{\bm u}^2} 
                        \right ],
\end{equation} 
with ${\bm \Omega} = {\bm \nabla}\times {\bm u}$.
The parameters $\go$ and $\gb$ in the above equations are defined in 
Eq.~(\ref{eqabc}).

Eq.~(\ref{Phieq}) can be re-written in terms of the metric
\begin{equation}
g^{ij} = (1+{\bm u}^2)(\delta^{ij}+u^iu^j).
\end{equation}
Then the equation looks like a Klein-Gordon equation in a curved
background with an ${\bm \Omega}\cdot {\bm \nabla}\Phi$ coupling 
and a source term. 

The solution of (\ref{Phieq}) can be inserted into the above
expression for ${\bm B}$ in Eq.~(\ref{Bsolnadv}) and that will be 
the steady state solution in the case of negligible diffusion.

The advection term is important for large velocities, in particular, 
for $v \gg \gb$, where $\gb$ is the parameter in the $\chi B$ effect.  
In this case, the second and third terms in the square bracket in
Eq.~(\ref{Bsolnadv}) are suppressed by powers of $\gamma_B$ and we 
get the order-of-magnitude estimate
\begin{equation}
B \approx - \frac{\go}{\gb} {\hat {\bm v}}\cdot {\bm \omega} ~ 
                               {\hat {\bm v}},
\end{equation}
where ${\hat {\bm v}}$ denotes a unit vector in the direction
of ${\bm v}$. The presence of kinetic helicity is crucial for this
estimate to be non-vanishing. 
Assuming helical flow with ${{\bm v}}\cdot {\bm \omega} \sim v\omega$, 
we get \begin{equation}
B  \sim - \frac{\go}{\gb} \omega,
\end{equation}
which coincides with the order of magnitude estimate in the
case of negligible advection, Eq.~(\ref{k0soln}).


\begin{thebibliography}{99}
 
\bibitem{Vachaspati:1991nm} 
  T.~Vachaspati,
  Phys.\ Lett.\ B {\bf 265}, 258 (1991).

\bibitem{Vachaspati:1994xc} 
  T.~Vachaspati,
  in the Proceedings of the NATO Workshop on "Electroweak Physics
  and the Early Universe", Sintra, Portugal (1994); Series B: Physics
  Vol. 338, Plenum Press, New York (1994);
  hep-ph/9405286.

\bibitem{Vachaspati:2008pi} 
  T.~Vachaspati,
  Phil.\ Trans.\ Roy.\ Soc.\ Lond.\ A {\bf 366}, 2915 (2008)
  [arXiv:0802.1533 [astro-ph]].

\bibitem{Kosowsky:1996yc}
A.~Kosowsky and A.~Loeb, Astrophys.J. {\bf 469}, 1 (1996).

\bibitem{Harari:1996ac}
D.~D.~Harari, J.~D.~Hayward, and M.~Zaldarriaga,
Phys.Rev. {\bf D55}, 1841 (1997).

\bibitem{Kosowsky:2004zh}
A.~Kosowsky, T.~Kahniashvili, G.~Lavrelashvili, and B.~Ratra
Phys. Rev. {\bf D71}, 043006 (2005).

\bibitem{Kahniashvili:2008hx}
T.~Kahniashvili, Y.~Maravin, and A.~Kosowsky,
Phys.Rev. {\bf D80}, 023009 (2009).

\bibitem{Pogosian:2011qv} 
  L.~Pogosian, A.~P.~S.~Yadav, Y.~-F.~Ng and T.~Vachaspati,
  Phys.\ Rev.\ D {\bf 84}, 043530 (2011)
  [Erratum-ibid.\ D {\bf 84}, 089903 (2011)]
  [arXiv:1106.1438 [astro-ph.CO]].

\bibitem{Cornwall:1997ms} 
  J.~M.~Cornwall,
  Phys.\ Rev.\ D {\bf 56}, 6146 (1997)
  [hep-th/9704022].

\bibitem{Vachaspati:2001nb} 
  T.~Vachaspati,
  Phys.\ Rev.\ Lett.\  {\bf 87}, 251302 (2001)
  [astro-ph/0101261].

\bibitem{Copi:2008he} 
  C.~J.~Copi, F.~Ferrer, T.~Vachaspati and A.~Achucarro,
  Phys.\ Rev.\ Lett.\  {\bf 101}, 171302 (2008)
  [arXiv:0801.3653 [astro-ph]].

\bibitem{Chu:2011tx}
  Y.~Z.~Chu, J.~B.~Dent and T.~Vachaspati,
  arXiv:1105.3744 [hep-th].
	
\bibitem{Caprini:2003vc}
 C.~Caprini, R.~Durrer and T.~Kahniashvili,
 Phys.\ Rev.\ D {\bf 69}, 063006 (2004)
 [astro-ph/0304556].

\bibitem{Kahniashvili:2005xe} 
  T.~Kahniashvili and B.~Ratra,
  Phys.\ Rev.\ D {\bf 71}, 103006 (2005)
  [astro-ph/0503709].

\bibitem{Kahniashvili:2005yp} 
  T.~Kahniashvili and T.~Vachaspati,
  Phys.\ Rev.\ D {\bf 73}, 063507 (2006)
  [astro-ph/0511373].
	
\bibitem{Vilenkin:1979ui} 
  A.~Vilenkin,
  Phys.\ Rev.\ D {\bf 20}, 1807 (1979).

\bibitem{Vilenkin:1980fu} 
  A.~Vilenkin,
  Phys.\ Rev.\ D {\bf 22}, 3080 (1980).

\bibitem{Vilenkin:1982pn} 
  A.~Vilenkin and D.~A.~Leahy,
  Astrophys.\ J.\  {\bf 254}, 77 (1982).
\bibitem{Kharzeev2006} 
  D.~E.~Kharzeev and A.~ Zhitnitsky, Nucl.\ Phys.\ A {\bf 797}, 67 (2007) [arXiv:0706.1026 [hep-ph]].

\bibitem{Kharzeev:2011vv} 
  D.~E.~Kharzeev,
  J.\ Phys.\ G G {\bf 38}, 124061 (2011)
  [arXiv:1107.4004 [hep-ph]].

\bibitem{KolbTurner}
E.W. Kolb and M.S. Turner, {\it The Early Universe} (Addison-Wesley, Reading,
MA, 1990).

\bibitem{Joyce:1997uy} 
  M.~Joyce and M.~E.~Shaposhnikov,
  Phys.\ Rev.\ Lett.\  {\bf 79}, 1193 (1997)
  [astro-ph/9703005].

\bibitem{Campbell:1992jd} 
  B.~A.~Campbell, S.~Davidson, J.~R.~Ellis and K.~A.~Olive,
  Phys.\ Lett.\ B {\bf 297}, 118 (1992)
  [hep-ph/9302221].

\bibitem{Pogosian:2001np} 
  L.~Pogosian, T.~Vachaspati and S.~Winitzki,
  Phys.\ Rev.\ D {\bf 65}, 083502 (2002)
  [astro-ph/0112536].

\bibitem{Boyarsky:2011uy} 
  A.~Boyarsky, J.~Frohlich and O.~Ruchayskiy,
  Phys.\ Rev.\ Lett.\  {\bf 108}, 031301 (2012)
  [arXiv:1109.3350 [astro-ph.CO]].

\bibitem{Turner:1987bw} 
  M.~S.~Turner and L.~M.~Widrow,
  Phys.\ Rev.\ D {\bf 37}, 2743 (1988).

\bibitem{Baym:1997gq}
 G.~Baym and H.~Heiselberg,
 Phys.\ Rev.\  D {\bf 56}, 5254 (1997)
 [arXiv:astro-ph/9704214].
	

\bibitem{Mukhanov:2005sc} 
  V.~Mukhanov,
  Cambridge, UK: Univ. Pr. (2005) 421 p
	
\bibitem{Jedamzik:2010cy}
 K.~Jedamzik and G.~Sigl,
 Phys.\ Rev.\ D {\bf 83}, 103005 (2011)
 [arXiv:1012.4794 [astro-ph.CO]].


\bibitem{Banerjee} 
  R.~Banerjee and K.~Jedamzik,
  Phys.\ Rev.\ D {\bf 70}, 123003 (2004)
  [astro-ph/0410032].

\bibitem{Campanelli} 
  L.~Campanelli,
  Phys.\ Rev.\ Lett.\  {\bf 98}, 251302 (2007)
  [arXiv:0705.2308 [astro-ph]].


\end{thebibliography}
\end{document}